\documentclass[prb,aps,amsmath,twocolumn,amsfonts,showpacs,superscriptaddress,floatfix]{revtex4}

\usepackage{bm}
\usepackage{graphicx}
\usepackage{color}

\def\be{\begin{equation}}
\def\ee{\end{equation}}

\def\Tr{\mathop{\rm Tr}}
\def\tr{\mathop{\rm tr}}
\def\sign{\mathop{\rm sign}}
\def\sn{\mathop{\rm sn}}
\def\var{\mathop{\rm var}}
\newcommand{\corr}[1]{\langle #1\rangle}
\newcommand{\ccorr}[1]{\langle\langle #1\rangle\rangle}

\def\eps{\varepsilon}

\def\PauliNambu{\tau}    
\def\PauliSpin{\sigma}   
\def\PauliDC{\hat\Sigma} 

\def\minigap{E_*}        

\begin{document}

\title{Mesoscopic fluctuations of the supercurrent
in diffusive Josephson junctions}

\author{Manuel Houzet}
\affiliation{Commissariat \`a l'\'Energie Atomique, DSM/DRFMC/SPSMS,
     38054 Grenoble, France}

\author{Mikhail A. Skvortsov}
\affiliation{Landau Institute for Theoretical Physics, Chernogolovka,
Moscow region, 142432 Russia}

\date{April 25, 2007} 


\begin{abstract}
We study mesoscopic fluctuations and weak localization correction
to the supercurrent in Josephson junctions with coherent diffusive
electron dynamics in the normal part.
Two kinds of junctions are considered: a chaotic dot coupled to
superconductors by tunnel barriers and a diffusive junction
with transparent normal--superconducting interfaces.
The amplitude of current fluctuations
and the weak localization correction to the average current are
calculated as functions of the ratio between the superconducting gap
and the electron dwell energy, temperature, and superconducting phase
difference across the junction.
Technically, fluctuations on top of the spatially inhomogeneous
proximity effect in the normal region are described by the
replicated version of the $\sigma$-model.
For the case of diffusive junctions with transparent
interfaces, the magnitude of mesoscopic fluctuations
of the critical current appears to be nearly 3 times larger
than the prediction of the previous theory
which did not take the proximity effect into account.
\end{abstract}

\maketitle

\section{Introduction}

At low temperatures, conductance of metals is due to electron
scattering on impurities.
The wave nature of electron motion reveals in a number of
quantum interference effects: the weak localization (WL) correction\cite{WL}
to the classical Drude conductance,
and universal sample-to-sample conductance fluctuations.\cite{ucf,washburn}
In mesoscopic samples whose size does not exceed the phase coherence
length, $L_\phi(T)$, the magnitude of conductance fluctuations
characterized by the root mean square (rms) $\delta G$ is independent
on the system size and degree of disorder and is of the order
of the conductance quantum $G_Q=e^2/\pi\hbar$.
Weak localization corrections and universal conductance fluctuations
have attracted considerable interest since the 80s' both from the
experimental and theoretical sides.\cite{MesoPhenomenaInSolids}

Two years after discovery of conductance fluctuations
in metals, Altshuler and Spivak\cite{altshuler87} applied
the same idea to fluctuations of the supercurrent
in Josephson junctions formed of a diffusive normal metal (N) placed
between two superconducting (S) leads. They considered the limit
of {\em long}\/ junctions, when the Thouless energy
$E_T=\hbar D/L^2$ ($D$ is the diffusion constant and $L$ is the
length of the junction) is much smaller than
the superconducting gap $\Delta$ in the leads.
In particular, it was found in Ref.~\onlinecite{altshuler87},
that for quasi-one-dimensional long wires,
mesoscopic fluctuations of the critical
current are characterized by the rms 
$\delta I_c = \sqrt{3\zeta(3)} \, eE_T/\pi\hbar
= 0.60 eE_T/\hbar$ at zero temperature.

The theory of Ref.~\onlinecite{altshuler87} was based on the
standard diagrammatic technique operating with soft diffusive
modes -- diffusons and Cooperons -- but Andreev reflection at the
NS interface\cite{Andreev} was described by the {\em linear}\/ phenomenological
boundary conditions on diffusive modes\cite{spivak-BC}.
However, the proximity effect in SNS systems is known to be essentially
nonperturbative at low energies,
where Cooper pairs penetrating from a superconductor strongly
modify electronic properties and open a minigap $\minigap$
in the normal region.\cite{minigap}
This strong perturbation of the metallic state can be described
only with the help of the full set of {\em nonlinear}\/ Usadel
equations\cite{Usadel} (in diffusive systems).
Thus, the treatment of Ref.~\onlinecite{altshuler87},
which effectively considered mesoscopic fluctuations
of the supercurrent on top of a uniform metallic state
without a minigap, is not accurate and should be
reconsidered taking the proximity effect into account
nonperturbatively.\cite{AS-comment}

Extensive studies of the proximity effect in SNS
systems\cite{kulik-omelyanchuk,ZaikinZharkov81,dubos,GKI04}
demonstrate that the critical Josephson current is related to the
minigap induced in the normal region: $I_c \sim G\minigap/e$,
where $G$ is the normal-state conductance of the junction.
On the other hand, the minigap can be roughly estimated as
$\minigap\sim\min(\Delta,E_\text{dwell})$, where $\Delta$
is the superconducting gap, and $E_\text{dwell}=\hbar/t_\text{dwell}$
is the energy associated with the typical dwell time $t_\text{dwell}$
of an electron in the normal region.
Since mesoscopic fluctuations for quantum dots and quasi-one-dimensional
systems are usually small in the parameter
$G_Q/G$, one can estimate the magnitude of mesoscopic fluctuations
as $\delta I_c \sim e\minigap/\hbar$.
For a long diffusive wire, $E_\text{dwell} \sim E_T$, indicating
that results of Ref.~\onlinecite{altshuler87}
are qualitatively correct.

The scattering matrix approach\cite{beenakker-RMT} has proven to be
a powerful tool for studying coherent electron transport
in mesoscopic conductors.
In the framework of this approach, an arbitrary
scatterer can be described by a set of transparencies
in each conduction channel.
The main transport characteristics of a mesoscopic conductor, such as
the conductance, shot noise power, conductance of the NS junction,
can be written as a linear statistics on the transmission eigenvalues,
$\sum_nf(T_n)$.\cite{beenakker-RMT}
Then, the corresponding weak localization corrections
and mesoscopic fluctuations can be determined from the knowledge
of the average density and the correlation function
of transmission eigenvalues. For a diffusive wire,
they were obtained in Refs.~\onlinecite{macedo,beenakker94}.
However, the Josephson current is generally {\em not}\/
a linear statistics on $T_n$'s.
It is a linear statistics only for a {\em short}\/
($\Delta\ll E_T$) wire:\cite{beenakker91}
$I(\chi)=(e\Delta/2\hbar)
\sum_n T_n\sin\chi/[1-T_n\sin^2(\chi/2)]^{1/2}$,
where $\chi$ is the superconducting phase difference.
In this limit one finds for the fluctuations of the critical
current at zero temperature:\cite{macedo,beenakker94}
$\delta I_c\simeq 0.30 (e\Delta/\hbar)$.
The crossover between short and long wires was also considered
numerically within the tight-binding model.\cite{takane}

Recently, the problem of mesoscopic fluctuations
of the supercurrent in short junctions with weakly transparent
NS interfaces was considered by Micklitz\cite{Micklitz}
who used the machinery of the supersymmetric nonlinear $\sigma$-model
\cite{Efetov-book}.
Within that approach, the average Josephson current is obtained
at the level of the saddle point which corresponds to the quasiclassical
Usadel equations\cite{AST}, while mesoscopic fluctuations can be
obtained from fluctuations around the saddle point.
We believe that the functional $\sigma$-model approach is a proper tool
for studying mesoscopic fluctuations of the Josephson current,
especially in the limit of long junctions where the multiple scattering
theory fails.

A couple of experiments with the use of a gated semiconductor
instead of a normal metallic part have been
reported.\cite{takayanagi,defranceschi}
Mesoscopic fluctuations were observed with varying the gate voltage.
In Ref.~\onlinecite{takayanagi}, mesoscopic fluctuations were
shown to follow precisely those of $G$, the latter being estimated
from $I(V)$ curves at a bias voltage larger than the
superconducting gap. On the other hand, $\delta I_c$ was
systematically found to be smaller than the theoretical estimate
from Ref.~\onlinecite{altshuler87}. A similar behavior was
reported in Ref.~\onlinecite{defranceschi}.

In this work, we present the derivation of mesoscopic fluctuations,
$\delta I(\chi)=[\corr{I^2(\chi)}-\corr{I(\chi)}^2]^{1/2}$,
and weak localization correction, $\Delta_\text{WL}I(\chi)$,
to the Josephson current in diffusive SNS junctions
within the replicated version of the nonlinear $\sigma$-model.
We ignore interaction effects in the normal metal
and assume zero magnetic field.
By expanding the action of the $\sigma$-model around its saddle
point, corresponding to an inhomogeneous solution of the Usadel
equations, we present mesoscopic fluctuations in terms of the
soft modes, analogous to Cooperons and diffusons in the normal
state\cite{Micklitz,AST}.
This approach allows us to follow the crossover between the
regimes of short and long wires.

In general, we verify 
that for the Josephson current through a chaotic dot
and quasi-one-dimensional wire,
$\delta I_c/I_c \sim - \Delta_\text{WL}I_c/I_c \sim G_Q/G$,
where $G$ is the normal-state conductance of the system,
and determine the exact coefficients in these relations
as functions of the junction length and temperature.
These coefficients are generally of the order of 1,
but in some cases (two tunnel barrier structures with
low barrier transparency) are additionally suppressed.

In particular, we find that the approach of Ref.~\onlinecite{altshuler87}
which does not take the proximity effect into account
systematically underestimates rms $I_c$ by the factor of order 3.
In the limit of long ($E_T\ll\Delta$)
quasi-one-dimensional junctions at zero temperature,
we obtain
\be
\label{AS-corrected}
  \delta I_c
  = 1.49 \frac{eE_T}{\hbar} ,
\ee
which is 2.5 times larger than the prediction of
Ref.~\onlinecite{altshuler87}.
Similarly, for wide ($W_x,W_y\gg L$) and long ($E_T\ll\Delta$)
three-dimensional junctions made of a metallic parallelepiped
of size $L\times W_y\times W_z$, we find
at $T=0$:
\be
\label{II-3D-long}
  \delta I_c
  = 2.0 \frac{e E_T}{\hbar} \sqrt{\frac{W_yW_z}{L^2}},
\ee
which is 2.8 times larger than the corresponding result of
Ref.~\onlinecite{altshuler87}.

The paper is organized as follows.
In Section \ref{sec:S-QD-S}, we consider the case of a chaotic diffusive
dot coupled to the superconducting leads through tunnel barriers.
This model allows us to introduce the method and is simple enough
to be solved analytically.
In Section \ref{sec:S-N-S}, we consider the case of a
diffusive wire with transparent NS interfaces.
Mesoscopic fluctuations of the critical current in two-dimensional (2D)
and three-dimensional (3D) geometries are calculated in Sec.~\ref{S:23D}.
We discuss the results in Conclusion.
Technical details are delegated to several Appendices.

\section{Superconductor--chaotic dot--superconductor junctions}
\label{sec:S-QD-S}

As a warm-up, in this section we consider a Josephson junction
formed of a ``zero-dimensional" chaotic dot in contact with two
superconducting reservoirs through tunnel junctions with
conductances $G_L$ and $G_R$. A possible realization of this system would be a
short diffusive wire, with the Thouless energy $E_T$ much
exceeding the superconducting gap $\Delta$ in the superconductors
and with the intra-dot conductance $G_N=2\pi G_Q E_T/\delta$ (where $\delta$ is
the mean level spacing in the dot) much exceeding $G_L$ and $G_R$.
As a consequence, the conductance of the structure
in its normal state is determined solely by the tunnel barriers:
$G=G_LG_R/(G_L+G_R)$.

The average Josephson current in such a system has been studied
by Aslamazov, Larkin, and Ovchinnikov\cite{Aslamazov}
within the tunnel Hamiltonian approach, by Kupriyanov and Lukichev\cite{KL}
with the help of the quasiclassical Usadel equations, and by
Brouwer and Beenakker\cite{brouwer} using the scattering approach\cite{beenakker91}
and Random Matrix Theory (RMT) \cite{beenakker-RMT}.
It was found that the amplitude of the
supercurrent is controlled by the ratio between the superconducting
gap $\Delta$ and the ``escape'' energy $E_g\sim G\delta/G_Q$.
The later energy scale is associated to the broadening of levels
in the dot due to coupling with the leads,
playing here the role of the dwell energy
$E_\text{dwell}$ defined in Introduction.

We shall rederive these results in the fermionic replica
$\sigma$-model language and then use this formalism to study
mesoscopic fluctuations of the Josephson current for arbitrary ratio
$\Delta/E_g$. A similar approach was very recently followed by
Micklitz\cite{Micklitz} who considered the effect of barrier
transparencies on the average supercurrent and its fluctuation
in the regime $\Delta \ll E_g$, within the framework of
the supersymmetric $\sigma$-model.

In Sec.~\ref{SS:sigma}, we introduce the replica $\sigma$-model for
this system. In Sec.~\ref{SS:QD-J}, we analyze its saddle point solution
and rederive the quasiclassical result for the Josephson current.
The fluctuation determinant is calculated in Sec.~\ref{SS:FluctDeterminant}.
It contains both weak localization corrections to the supercurrent
and its mesoscopic fluctuations, which are analyzed in Secs.~\ref{SS:QD-WL}
and \ref{SS:QD-Mesofluct}.
The results are summarized in Sec.~\ref{SS:QD-results}.

\subsection{Replica $\bm\sigma$-model for a chaotic dot}
\label{SS:sigma}

The equilibrium supercurrent which flows in a Josephson junction
can be obtained from the free energy ${\cal F}=-kT\ln Z$ of the
system at temperature $T$:
\be
\label{eq:courant-energie}
   I(\chi)=\frac{2 e}{\hbar}\frac{d}{d\chi} {\cal F}(\chi),
\ee
where $\chi$ is the superconducting phase difference between the leads.
Disorder-averaging is performed in a standard way
using the replica trick,\cite{replicas}
\be
\label{eq:log-Z-average}
   \corr{{\cal F(\chi)}}
   = -kT\lim_{n\rightarrow 0} \frac{\corr{Z_\chi^n}-1}{n},
\ee
where $\corr{Z_\chi^n}$ can be evaluated as a functional
integral within the fermionic replica
$\sigma$-model:\cite{Wegner,ELK1980,Fin}
\be
\label{eq:action-sigma-model}
   \corr{Z_\chi^n}
   =
   \int {\cal D} Q \, e^{-S[Q]}.
\ee
The nonlinear $\sigma$-model is a field theory
formulated in terms of the matrix field $Q$ acting
in the direct product of
the replica space of dimension $n$,
infinite Matsubara energy space,
two-dimensional Gorkov-Nambu space (Pauli matrices $\PauliNambu_i$),
and two-dimensional spin space (Pauli matrices $\PauliSpin_i$).
The $Q$ matrix is subject to the nonlinear constraint
$Q^2=1$ and obeys the charge conjugation symmetry:
\be
   Q = \overline{Q}
   \equiv \PauliNambu_1 \PauliSpin_2 Q^T \PauliSpin_2 \PauliNambu_1 ,
\label{ChargeConjSym}
\ee
where $Q^T$ stands for the full matrix transposition.
The condition (\ref{ChargeConjSym}) is related to simultaneous
introduction of the Gorkov-Nambu and spin spaces, which renders the vectors
$\Psi=(\psi_\uparrow, \psi_\downarrow^*, \psi_\downarrow, -\psi_\uparrow^*)^T$
and $\Psi^*$ be linearly dependent.
The functional integration in $Q$ is performed over an appropriate
real submanifold of the complex manifold defined by the constraints
$Q^2=1$ and $Q=\overline{Q}$.

The action of the $\sigma$-model for a chaotic dot coupled
to the superconducting terminals via tunnel junctions
is given by \cite{oreg}
\be
\label{eq:action}
   S[Q]
   =
   -
   \frac{\pi}{2\delta}
   \tr
   \biggl(
   \epsilon\PauliNambu_3 Q
   + \sum_{i=L,R} \frac{ G_i\delta}{4\pi G_Q} Q_i Q
   \biggr).
\ee
Here, $\delta=(\nu V)^{-1}$ is the mean level spacing in the dot
($V$ is the dot's volume and $\nu$ is the single-particle density
of states at the Fermi energy per one spin projection),
$\epsilon$ is the fermionic
Matsubara energy, and the trace is taken over all spaces
of the $Q$ matrix.

In the superconducting reservoirs with the order parameters
$\Delta e^{-i\chi/2}$ (left) and $\Delta e^{i\chi/2}$ (right),
the matrices $Q_i$ ($i=L$, $R$) are unit matrices in the replica and spin spaces,
being diagonal in the energy space with the matrix elements:
\be
\label{QS}
   Q_{L,R}
   =
   \left(
     \PauliNambu_1\cos \frac{\chi}{2} \pm \PauliNambu_2\sin\frac{\chi}{2}
   \right)
   \sin\theta_s
   +
   \PauliNambu_3
   \cos\theta_s,
\ee
where
$\cos\theta_s=\cos\theta_s(\epsilon)=\epsilon/\sqrt{\epsilon^2+\Delta^2}$.

Equation (\ref{QS}) is often referred to as ``the rigid boundary condition''.
It corresponds to neglecting the inverse proximity effect as well as
depairing effect in the leads due to a finite current density (see, e.g.,
review~\onlinecite{GKI04}).

\subsection{Saddle point: average Josephson current}
\label{SS:QD-J}

We start the analysis of the $\sigma$-model (\ref{eq:action})
with the saddle-point approximation, which amounts to neglecting
mesoscopic fluctuations and weak-localization corrections.
The matrix $Q_0$ which extremizes the action
solves the saddle-point equation:
\be
\label{eq:usadel-dot}
   \biggl[
   \epsilon \PauliNambu_3
   + \sum_{i=L,R} \frac{G_i\delta}{4\pi G_Q} Q_i
, Q_0 \biggr]
   = 0.
\ee
Equation (\ref{eq:usadel-dot}) is nothing but the Usadel equation
for the quasiclassical Green's function for a chaotic dot supplied by the
Kupriyanov-Lukichev boundary conditions\cite{KL} at the tunnel interface.
This equation can also be obtained with the help of
Nazarov's ``circuit theory''.\cite{nazarov}

\begin{figure}
\includegraphics{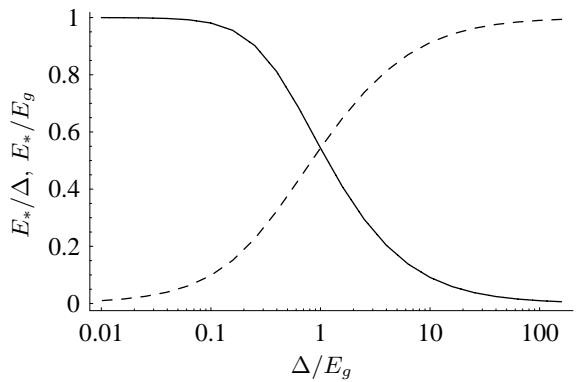}
\caption{The minigap $\minigap$ vs.\ $\Delta/E_g$
in a superconductor--quantum dot--superconductor Josephson
junction with symmetric tunnel barriers. The two curves
correspond to $\minigap$ in units of $\Delta$ (solid line)
and in units of $E_g$ (dashed line).
}
\label{F:qd-minigap}
\end{figure}

The solution of Eq.~(\ref{eq:usadel-dot})
proportional to the unit matrix in the replica and spin spaces
can be easily found:
\begin{equation}
\label{Q0}
Q_0
=
(\PauliNambu_1\cos \phi-\PauliNambu_2\sin\phi)\sin\theta+\PauliNambu_3\cos\theta ,
\end{equation}
where
\begin{subequations}
\label{angles}
\begin{gather}
   \tan\theta(\epsilon)
   =
   \frac{\Delta \, E_g(\chi)}{\epsilon (\sqrt{\epsilon^2+\Delta^2}+E_g)} ,
\\
   \tan\phi
   =
   \frac{G_R-G_L}{G_L+G_R}\tan\chi,
\end{gather}
\end{subequations}
and
\be
   E_g(\chi)
   =
   \frac{\delta}{4\pi G_Q}
   \sqrt{G_L^2+G_R^2+2G_LG_R\cos\chi} ,
\label{Eg(chi)}
\ee
with $E_g\equiv E_g(0)$.
The pole of $Q_0$ located at imaginary
$\epsilon$ is related to the minigap $\minigap(\chi)$
in the density of states of the normal island.\cite{minigap}
In the limiting cases,
\be
   \minigap(\chi)
   =
   \begin{cases}
     E_g(\chi) , & \Delta\gg E_g , \\
     (\Delta/E_g) E_g(\chi), & \Delta\ll E_g ,
   \end{cases}
\ee
while in the intermediate region, $\Delta\sim E_g$, the dependence
of $\minigap$ on $\chi$ is more complicated.
In what follows, we will denote $\minigap=\minigap(0)$.
Roughly speaking, $\minigap\approx\min(\Delta,E_g)$, see
Fig.~\ref{F:qd-minigap}.

The action at the saddle point is given by $nS_0$,
where $n$ is the number of replicas and
\be
\label{eq:action-saddle}
   S_0(\chi)
   =
   -\frac{2\pi}{\delta}
   \sum_\epsilon
   \omega_\epsilon(\chi) ,
\ee
with the summation over the fermion Matsubara energies
$\epsilon_p=\pi(2p+1)kT$, and
\be
   \omega_\epsilon(\chi)
   =
   \frac{\sqrt{
     \epsilon^2
     (\sqrt{\epsilon^2+\Delta^2}
     +E_g)^2
     +\Delta^2 E_g(\chi)^2}}
     {\sqrt{\epsilon^2+\Delta^2}} .
\label{omega-def}
\ee
In the leading order in $G/G_Q\gg 1$, one can neglect weak localization
and mesoscopic fluctuations effects (they will be studied in the next
subsections). In this approximation, equivalent to the standard
quasiclassical analysis, the Gaussian integral near the saddle point
yields unity, and the average Josephson current can now be obtained from
Eqs.~(\ref{eq:courant-energie})--(\ref{eq:action-sigma-model}) as
$\corr{I(\chi)}_0 = (2ekT/\hbar) \partial S_0(\chi)/\partial\chi$,
yielding
\begin{equation}
   \langle I(\chi) \rangle_0
   =
   \frac{\pi kT}{e}
   GE_g
   \sum_\epsilon
   \frac{\Delta^2\sin\chi}{(\epsilon^2+\Delta^2)\omega_\epsilon(\chi)}.
\label{eq:I-dot}
\end{equation}

The result (\ref{eq:I-dot}) is certainly not new.
It had been obtained previously by a number
of authors\cite{Aslamazov,KL,brouwer,Micklitz}.
Here we simply rederive this result for completeness.

\begin{figure}
\includegraphics{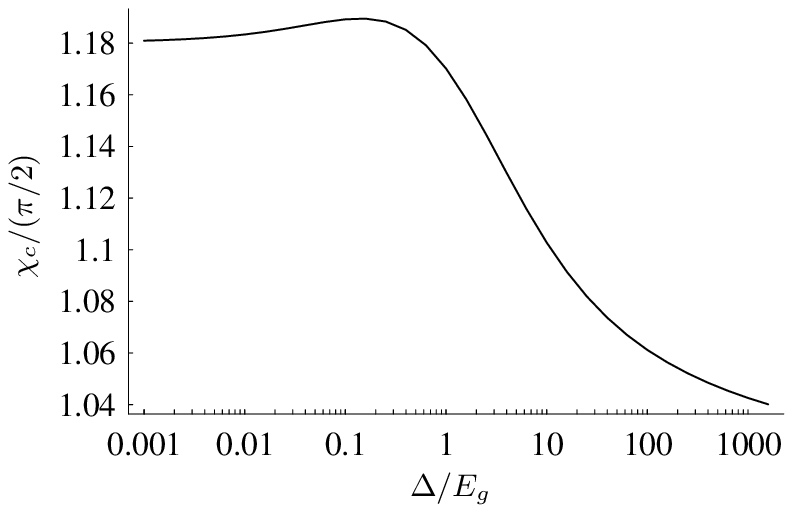}
\includegraphics{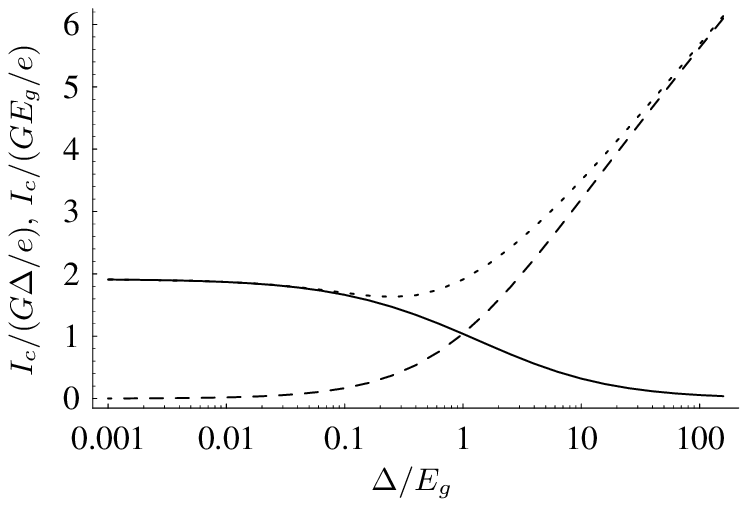}
\caption{Quasiclassical results for a superconductor--quantum dot--superconductor
Josephson junction with symmetric tunnel barriers:
(a) the critical phase $\chi_c$ (in units of $\pi/2$)
vs.\ $\Delta/E_g$ at zero temperature;
(b) the critical current $I_c$ (solid line: in units of $G \Delta/e$,
dashed line: in units of  $G E_g/e$, dotted line: in units of
$G\minigap/e$) vs.\ $\Delta/E_g$
at zero temperature.}
\label{F:Ic-chic-dot}
\end{figure}

\subsubsection*{Symmetric junction}

The general expression (\ref{eq:I-dot}) simplifies
for symmetric barriers: $G_L=G_R=2G$.
In this case, $E_g(\chi)=E_g\, |{\cos(\chi/2)}|$ and $E_g=G\delta/\pi G_Q$.

{\em At zero temperature} (more precisely, $kT\ll \minigap$),
the average Josephson current is controlled
by the ratio $\Delta/E_g$.

When $\Delta\ll E_g$, the Josephson relation is not sinusoidal:
\be
   \langle I(\chi) \rangle_0
   =
   \frac{G\Delta}{e}
   \sin\chi \,
   {\mathbf{K}} \left(\sin\frac{\chi}{2}\right) ,
\ee
where $\mathbf{K}(x)=F(\pi/2,x)$ is the full elliptic integral
of the first kind defined as in Ref.~\onlinecite{GradsteinRyzhik}.
The critical current $I_c\simeq 1.92 G\Delta/e$
is achieved at a phase difference $\chi_c\simeq 1.18(\pi/2)$.

In the opposite limit, $\Delta\gg E_g$,
the Josephson relation is close to sinusoidal:
\begin{equation}
\langle I(\chi) \rangle_0
\simeq
\frac{GE_g}{e}
\sin\chi
\ln\left(\frac{2\Delta}{E_g(\chi)}\right) ,
\end{equation}
with the
critical current $I_c\simeq(GE_g/e)\ln(2\Delta/E_g)$
at a phase difference $\chi_c\simeq\pi/2$.

The crossover for the critical current and critical phase at zero
temperature and arbitrary relation between $\Delta$ and $E_g$ is
illustrated in Fig.~\ref{F:Ic-chic-dot}.

{\em At temperature close to the critical temperature of the leads}, $T_c$,
the superconducting gap vanishes as $\Delta(T)\propto k[T_c(T_c-T)]^{1/2}$.
Then for relevant energies $\epsilon \gtrsim kT\gg \Delta(T)$,
Eq.~(\ref{eq:I-dot}) can be simplified, yielding a
sinusoidal Josephson relation with the critical current
\be
   I_c
   =
   \begin{cases}
     (\pi G \Delta^2/4ekT_c) , & \Delta\ll E_g , \\
     7\zeta(3)GE_g\Delta^2/4e\pi^2k^2T_c^2, & \Delta\gg E_g ,
   \end{cases}
\ee
where $\zeta(3)\approx1.202$ is the Riemann zeta function.

{\em At intermediate temperatures}, $E_g \ll kT \ll kT_c$,
we find $I_c = (GE_g/e)\ln(2\gamma\Delta/\pi kT)$,
where $\gamma=e^C=1.781\dots$ is the Euler constant,
in agreement with Ref.~\onlinecite{Aslamazov,KL,brouwer}.

\subsection{Gaussian fluctuations near the saddle point}
\label{SS:FluctDeterminant}

In this subsection we take Gaussian fluctuations near the saddle
point into account. Since we will be interested in mesoscopic
fluctuations of the Josephson current (see
Sec.~\ref{SS:QD-Mesofluct}), we will have to consider the average
$\corr{Z_{\chi_1}^{n_1}Z_{\chi_2}^{n_2}}$ of two partition
functions calculated at different superconducting phases $\chi_1$
and $\chi_2$. This average can also be expressed in terms of the
$\sigma$-model:
\be
\label{eq:prod-partition}
   \corr{Z_{\chi_1}^{n_1}Z_{\chi_2}^{n_2}}
   = \int {\cal D} Q \, e^{-S[Q]},
\ee
where the only difference with the $\sigma$-model
described in Sec.~\ref{SS:sigma} is that now $Q$ becomes an
$(n_1+n_2)\times (n_1+n_2)$ matrix in the replica space.
Correspondingly, the superconducting $Q$-matrices in the terminals
should be modified.
Now they are diagonal in the replica space,
having the superconducting phase difference
$\chi_1$ (resp.~$\chi_2$) in the $n_1$ first (resp.~$n_2$ last) replicas.
With these modification,
the action of the $\sigma$-model has the same form (\ref{eq:action}).

At the saddle-point, the matrix $Q_0$ extremizing the action is
diagonal in the energy and replica spaces. Its diagonal elements are
given by Eqs.~(\ref{Q0}) and (\ref{angles}), where the phase
$\chi$ is set to $\chi_1$ (resp.~$\chi_2$) if $1\leq a\leq n_1$
(resp.\ $n_1<a \leq n_1+n_2$), where $a$ is a replica index.

In order to study fluctuations near this saddle point we write
matrices close to $Q_0$ as
\be
\label{Q-W}
   Q = U^\dagger \Lambda (1+W+W^2/2+\dots) U.
\ee
Here, the matrices $\Lambda$ and $U$ should be chosen
in such a way that in the absence of fluctuations, at $W=0$,
Eq.~(\ref{Q-W}) reduces to $U^\dagger\Lambda U = Q_0$.
As usually, one has to require
\begin{subequations}
\be
\label{Lambda-W}
   \{\Lambda,W\} = 0 ,
\ee
and impose the constraint following from
Eq.~(\ref{ChargeConjSym}):
\be
   \overline W = - W ,
\label{W-bar}
\ee
and the requirement of convergency of the $\sigma$-model
on the perturbative level:
\be
\label{W-dagger}
   W^\dagger = - W .
\ee
\end{subequations}

The form of the parametrization (\ref{Q-W}) is standard,
while the choice of the matrix $\Lambda$ is a matter of convenience.
A possible choice could be the metallic saddle point,\cite{Fin}
$\Lambda_M=\tau_3\sign (\epsilon)$  [the limit of Eq.~(\ref{QS})
at $\epsilon\gg\Delta$].
Here we adopt an alternative choice
proposed by Ostrovsky and Feigel'man\cite{superconductingSP}:
\be
\label{Lambda-choice}
   \Lambda = \PauliNambu_1 ,
\ee
corresponding to the superconducting saddle point~(\ref{QS})
at zero energy and $\chi=0$.
With this choice of $\Lambda$, the unitary matrix $U$
in Eq.~(\ref{Q-W}) is given by
\be
   U
   =
   e^{-i\PauliNambu_2(\pi/4-\theta/2)}
   e^{-i\PauliNambu_3(\phi/2)} .
\label{U}
\ee
The choice of the parametrization with $\Lambda=\PauliNambu_1$
has two technical advantages: (i) the solution
of the constraint (\ref{Lambda-W}) is independent on energy,
and (ii) the constraint (\ref{W-bar}) can be easily resolved.

A general parametrization of the matrix $W$ satisfying the constraint
(\ref{Lambda-W}) is given by:
\be
   W_{mn}=\PauliNambu_3 \hat d_{mn}+\PauliNambu_2 \hat c_{mn},
\label{eq:W}
\ee
where $m=(\epsilon,a)$ and $n=(\epsilon',b)$ encode both the energy
and replica indices, and $\hat d_{mn}$ and $\hat c_{mn}$
are $2\times2$ matrices in the spin space which can be expanded
in the Pauli matrices as
\be
   \hat d = d_0 + \mathbf{d} \bm{\PauliSpin} ,
   \qquad
   \hat c = c_0 + \mathbf{c} \bm{\PauliSpin} .
\ee
The variables $d_0$ and $c_0$ ($\mathbf{d}$ and $\mathbf{c}$)
will be referred to as singlet (triplet) modes.
They play the same role as diffusons and Cooperons in a normal metal,
describing soft diffusive excitations on top of an inhomogeneous
proximity-induced state. Note that contrary to diffusive modes
in a normal metal, these $d$- and $c$-modes are generally coupled
in the presence of a supercurrent in the normal region,
cf.\ Sec.~\ref{sec:S-N-S}.

Equations (\ref{W-bar}) and (\ref{W-dagger}) yield
\begin{subequations}
\label{dc-sym}
\begin{gather}
\label{ddd}
   d_0 = d_0^T = - d_0^\dagger ,
   \qquad
   \mathbf{d} = - \mathbf{d}^T = - \mathbf{d}^\dagger ,
\\
\label{ccc}
   c_0 = - c_0^T = - c_0^\dagger ,
   \qquad
   \mathbf{c} = \mathbf{c}^T = - \mathbf{c}^\dagger .
\end{gather}
\end{subequations}
Here, transposition acts in the replica and energy spaces.
In terms of the matrix elements, Eqs.~(\ref{ddd}), e.g., read:
\begin{subequations}
\label{d-sym-elements}
\begin{gather}
   (d_0)^{ab}_{\epsilon,\epsilon'}
= (d_0)^{ba}_{-\epsilon',-\epsilon}
= - (d_0^*)^{ba}_{\epsilon',\epsilon} ,
\\
   (d_i)^{ab}_{\epsilon,\epsilon'}
= - (d_i)^{ba}_{-\epsilon',-\epsilon}
= - (d_i^*)^{ba}_{\epsilon',\epsilon} .
\end{gather}
\end{subequations}
Independent integration variables
for the singlet $d_0$ mode can be chosen, e.g., as:
\begin{align}
   &(d_0)^{ab}_{\epsilon,\epsilon'} \in \mathbb{C}, \qquad \text{if $a>b$, $\epsilon>0$} ,
\nonumber
\\
   &(d_0)^{aa}_{\epsilon,\epsilon'} \in \mathbb{C}, \qquad \text{if $\epsilon>|\epsilon'|>0$} ,
\nonumber
\\
   &(d_0)^{aa}_{\epsilon,-\epsilon} \in \mathbb{C}, \qquad \text{if $\epsilon>0$} ,
\nonumber
\\
   &(d_0)^{aa}_{\epsilon,\epsilon} \in i \mathbb{R}, \qquad \text{if $\epsilon>0$} ,
\nonumber
\end{align}
and analogously for the triplet $c_{i=1,2,3}$.
Independent integration variables for the triplet $d_{i=1,2,3}$
modes can be chosen, e.g., as:
\begin{align}
   &(d_i)^{ab}_{\epsilon,\epsilon'} \in \mathbb{C}, \qquad \text{if $a>b$, $\epsilon>0$} ,
\nonumber
\\
   &(d_i)^{aa}_{\epsilon,\epsilon'} \in \mathbb{C}, \qquad \text{if $\epsilon>|\epsilon'|>0$} ,
\nonumber
\\
   &(d_i)^{aa}_{\epsilon,\epsilon} \in i \mathbb{R}, \qquad \text{if $\epsilon>0$} ,
\nonumber
\end{align}
and analogously for the singlet $c_0$.

Expanding the action in the Gaussian approximation
over fluctuations near the saddle point, one finds in general:
\begin{equation}
\label{S2-QD}
   S^{(2)}
   =
   \frac{\pi}{\delta}
   \sum_{mn}
   \sum_{i=0}^3
   \begin{pmatrix} d^*_{i} & c^*_{i} \end{pmatrix}_{mn}
   \hat{A}_{mn}
   \begin{pmatrix} d_{i}\\c_{i} \end{pmatrix}_{mn} ,
\end{equation}
where $\hat{A}_{mn}$ is a symmetric [it can be symmetrized using
relations (\ref{dc-sym})] matrix in the $(d,c)$-space
with the simple block structure in the replica space:
\begin{equation}
   \hat{A}_{\epsilon\epsilon'}^{ab}
   =
   \begin{cases}
   A_{\epsilon\epsilon'}^{\chi_1\chi_1} ,
     & \text{if $a,b\leq  n_1$},
   \\
   A_{\epsilon\epsilon'}^{\chi_1\chi_2} ,
     & \text{if $a\leq  n_1 <b$},
   \\
   A_{\epsilon\epsilon'}^{\chi_2\chi_1} ,
     & \text{if $b\leq n_1 < a$},
   \\
   A_{\epsilon\epsilon'}^{\chi_2\chi_2} ,
     & \text{if $n_1 < a,b$}.
   \end{cases}
\end{equation}
The matrix $\hat A$ does not depend on the spin index $i$ since
the spin in conserved. Such a dependence will arise if one takes
magnetic impurities or spin-orbit interaction in the normal region
into account.

Due to the absence of the $(d_i)_{\epsilon,-\epsilon}^{aa}$
and $(c_0)_{\epsilon,-\epsilon}^{aa}$ modes,
the matrix $\hat A_{\epsilon,-\epsilon}^{aa}$
should be diagonal in the $(d,c)$-space:
\be
   A_{\epsilon,-\epsilon}^{\chi\chi}
   =
   \begin{pmatrix}
     \bigl( A_{\epsilon,-\epsilon}^{\chi\chi} \bigr)^{dd} & 0 \\
     0 & \bigl( A_{\epsilon,-\epsilon}^{\chi\chi} \bigr)^{cc}
   \end{pmatrix}
   .
\ee
We will see below that this is indeed the case
[cf.\ Eqs.~(\ref{eq:A-dot}) and (\ref{A})].

Integration over independent variables of the $d$ and $c$ modes
gives the fluctuation determinant:
\be
\label{ZZ}
   \corr{Z_{\chi_1}^{n_1}Z_{\chi_2}^{n_2}}
   =
   {\cal M} \,
   e^{-n_1 \tilde S_0(\chi_1) - n_2 \tilde S_0(\chi_2)} ,
\ee
where $\tilde S_0(\chi)$ contains the WL correction:
\be \label{eq:WL}
   \tilde S_0(\chi)
   = S_0(\chi)
   - \frac12 \sum_{\epsilon}
     \tr \ln
     \frac
       {(A^{\chi\chi}_{\epsilon,-\epsilon})^{dd}}
       {(A^{\chi\chi}_{\epsilon,-\epsilon})^{cc}}
\ee
[we write this expression in the most general way assuming that
$(A^{\chi\chi}_{\epsilon,-\epsilon})^{dd}$
and $(A^{\chi\chi}_{\epsilon,-\epsilon})^{cc}$ might be operators,
as in Sec.~\ref{sec:S-N-S}; for a chaotic dot considered in this section,
the trace in Eq.~(\ref{eq:WL}) can be omitted],
whereas the prefactor ${\cal M}$ accounts for mesoscopic fluctuations:
\be
   {\cal M}
   =
   \prod_{\epsilon,\epsilon'}
   \frac{1}{
   \left(\det A_{\epsilon\epsilon'}^{\chi_1\chi_1}\right)^{n_1^2}
   \left(\det A_{\epsilon\epsilon'}^{\chi_1\chi_2}\right)^{2n_1 n_2}
   \left(\det A_{\epsilon\epsilon'}^{\chi_2\chi_2}\right)^{n_2^2}} .
\label{eq:MF}
\ee
In Eq.~(\ref{eq:MF}),
the product over $\epsilon$ and $\epsilon'$ should be taken
over all Matsubara energies and we have omitted the factors which
are equal to 1 in the replica limit $n_{1,2}\to0$
and do not depend on $\chi_{1,2}$.

For a superconductor--quantum dot--superconductor junction considered
in this Section, the matrix $\hat A$ can be easily found by expanding
the action (\ref{eq:action}) in $W$ with the help of
Eqs.~(\ref{Q-W})--(\ref{eq:W}):
\be
\label{eq:A-dot}
   A^{\chi\chi'}_{\epsilon\epsilon'}
   = \frac{\omega_\epsilon(\chi)+\omega_{\epsilon'}(\chi')}2 \, \PauliDC_0 ,
\ee
where $\omega_\epsilon(\chi)$ is defined in Eq.~(\ref{omega-def}),
and $\PauliDC_0$ is the identity matrix in the $(d,c)$-space.

The weak localization correction to the Josephson current
and its mesoscopic fluctuations are discussed in the next Subsections.

\subsection{Weak localization correction}
\label{SS:QD-WL}

The weak localization correction to the Josephson current,
$\Delta_{\text{WL}} I(\chi) \equiv \corr{I(\chi)} - \corr{I(\chi)}_0$,
can be found with the help of Eqs.~(\ref{eq:courant-energie}),
(\ref{eq:log-Z-average}), (\ref{ZZ}) and (\ref{eq:WL}):
\be
\label{eq:IWL}
   \Delta_{\text{WL}} I(\chi)
   =
   - \frac{e k T}{\hbar}
   \frac{\partial}{\partial\chi}
   \sum_{\epsilon} \tr \ln
     \frac
       {(A^{\chi\chi}_{\epsilon,-\epsilon})^{dd}}
       {(A^{\chi\chi}_{\epsilon,-\epsilon})^{cc}} .
\ee

Since for a superconductor--quantum dot--superconductor junction,
the matrix $\hat{A}$ [see Eq.~(\ref{eq:A-dot})] acts as a unit matrix
in the $(d,c)$-space, there is no weak localization correction
to the Josephson current.

Note that there is no weak localization correction
to the conductance of a normal metal--quantum dot--normal metal junction
in the lowest order in the tunnel barriers either \cite{beenakker-RMT,iida}.

\subsection{Mesoscopic fluctuations}
\label{SS:QD-Mesofluct}

With the help of the replica trick and relation (\ref{eq:courant-energie}),
the current-current correlation function
can be expressed as
\be
\label{eq:fluct-courant}
   \langle I(\chi_1) I(\chi_2) \rangle
   =
   \left(\frac{2 e k T}{\hbar}\right)^2
   \frac{\partial^2}{\partial\chi_1\partial\chi_2}
   \lim_{n_1,n_2\rightarrow 0 }
   \frac{\langle Z_{\chi_1}^{n_1}Z_{\chi_2}^{n_2} \rangle}{n_1n_2} .
\ee
Making use of Eqs.~(\ref{ZZ}) and (\ref{eq:MF}), we
get the general expression for the cumulant
$\ccorr{I(\chi_1) I(\chi_2)} \equiv \corr{I(\chi_1)
I(\chi_2)} - \corr{I(\chi_1)} \corr{I(\chi_2)}$:
\be
\label{eq:II}
\ccorr{I(\chi_1) I(\chi_2)}
=
-8\left(\frac{e k T}{\hbar}\right)^2
\frac{\partial^2}{\partial\chi_1\partial\chi_2}
\sum_{\epsilon\epsilon'} \tr \ln A^{\chi_1\chi_2}_{\epsilon\epsilon'}.
\ee

Since $A^{\chi_1\chi_2}_{\epsilon\epsilon'}$ given by Eq.~(\ref{eq:A-dot})
appears to be diagonal in the $(d,c)$-space,
evaluation of Eq.~(\ref{eq:II})
for the quantum dot geometry is trivial and we get:
\begin{multline}
\label{QD-II-cumulant}
   \ccorr{I(\chi_1) I(\chi_2)}
\\ {}
   =
   \left(\frac{4 e k T}{\hbar}\right)^2
   \sum_{\epsilon\epsilon'}
   \frac{1}{[\omega_\epsilon(\chi_1)+\omega_{\epsilon'}(\chi_2)]^2}
   \frac{\partial \omega_\epsilon}{\partial\chi_1}
   \frac{\partial \omega_{\epsilon'}}{\partial\chi_2}.
\end{multline}
In the limits $T=0$ and $\Delta\ll E_g$, Eq.~(\ref{QD-II-cumulant})
had been recently derived by Micklitz\cite{Micklitz} using the
supersymmetric $\sigma$-model approach.

As shown in Ref.~\onlinecite{Beenakker93},
mesoscopic fluctuations of the {\em critical}\/ current, $\delta I_c$,
can be obtained as $\delta I_c = \delta I(\chi_c)$.

\begin{figure}
\includegraphics{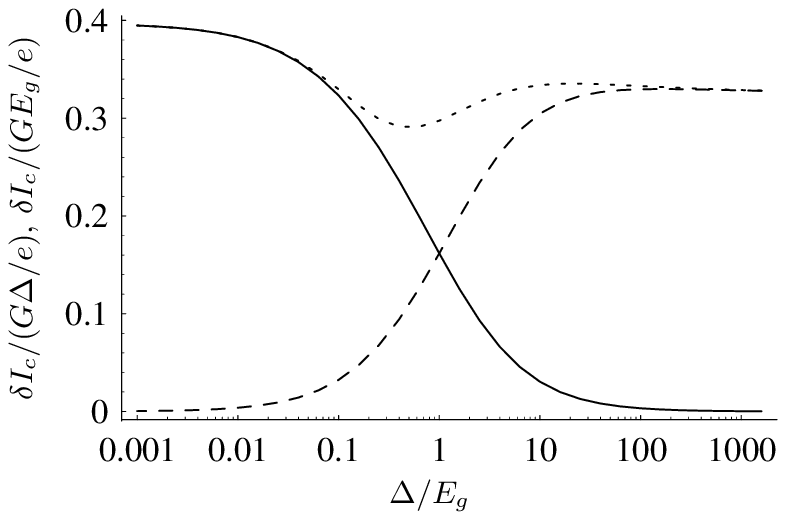}
\includegraphics{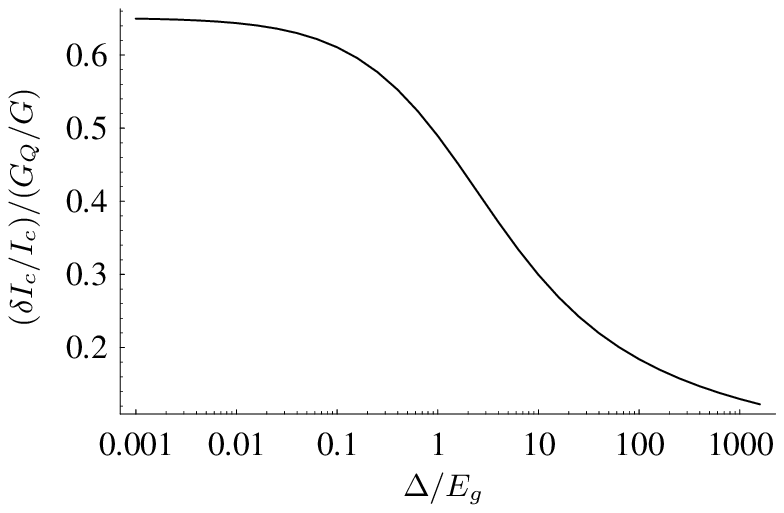}
\caption{(a) Mesoscopic fluctuations of the critical current,
$\delta I_c = (\mathop{\rm var}I_c)^{1/2}$
(solid line: in units of $e\Delta/\hbar$,
dashed line: in units of $e E_g/\hbar$,
dotted line: in units of $e\minigap/\hbar$)
vs.\ $\Delta/E_g$ at zero temperature.
(b) The ratio $G\delta I_c/G_QI_c$ vs.\ $\Delta/E_g$ at zero temperature.
}
\label{F:dIc-dot}
\end{figure}

\begin{table*}
\caption{Summary of results for a quantum dot contacted symmetrically
to superconducting leads ($G_L=G_R=2G$).}
\label{tab:dot}
\begin{ruledtabular}
\begin{tabular}{ccccccc}
     & $\chi_c/(\pi/2)$ & $e I_c/G$ & $\Delta_{\text{WL}} I(\chi)$ & $\hbar\delta I_c/e$ & $g\delta I_c/I_c$  \\
\hline
$\Delta\ll E_g,\, T=0$ & $1.18$ & $1.92\Delta{}^\text{[\onlinecite{KL,brouwer}]}$ & 0 & $0.396\Delta{}^\text{[\onlinecite{Micklitz}]}$ & 0.648\\
$E_g\ll \Delta,\, T=0$ & $1$ & $E_g\ln(2\Delta/E_g){}^\text{[\onlinecite{brouwer}]}$ & 0 & $E_g/\pi$ & $1/\ln(2\Delta/E_g)$ \\
arbitrary $E_g/\Delta$, $T=0$ & Fig.~\ref{F:Ic-chic-dot}(a) & Fig.~\ref{F:Ic-chic-dot}(b) & 0 & Fig.~\ref{F:dIc-dot}(a) & Fig.~\ref{F:dIc-dot}(b) \\
$kT \lesssim kT_c\ll E_g$ & 1 & $\pi \Delta^2(T)/4kT_c{}^\text{[\onlinecite{KL}]}$ & 0 & $\Delta^2(T)/8kT_c$ &$1/2$  \\
$E_g\ll kT \lesssim kT_c$  & 1 &$0.213 E_g\Delta^2(T)/(kT_c)^2{}^\text{ [\onlinecite{KL}]}$ & 0 & $0.010 \Delta^2(T) E_g^2/(kT_c)^3$&$0.15E_g/kT_c$ \\
$E_g\ll kT \ll kT_c$ & $1$ & $E_g\ln(2\gamma\Delta/\pi kT){}^\text{[\onlinecite{Aslamazov,KL,brouwer}]}$ & 0 & $0.12E_g^2/kT$ & $0.38E_g/[kT\ln(2\gamma\Delta/\pi kT)]$ \\
\end{tabular}
\end{ruledtabular}
\vskip 3pt
\footnotesize
Here, $g=G/G_Q$, $\Delta(T)=[(8\pi^2/7\zeta(3))k^2T_c(T_c-T)]^{1/2}$
and $\gamma=1.781\dots$
\hfill{}
\end{table*}

\subsubsection*{Symmetric junction}

We consider now symmetric junctions.

{\em At zero temperature}, we find:
\be
   \delta I_c
   =
   \begin{cases}
     0.396 (e \Delta/\hbar), & \text{if $\Delta\ll E_g$}, \\
     e E_g/\pi\hbar, & \text{if $E_g\ll\Delta$},
   \end{cases}
\ee
while the result for an arbitrary ratio $\Delta/E_g$ is plotted
in Fig.~\ref{F:dIc-dot}.

\begin{figure}
\includegraphics{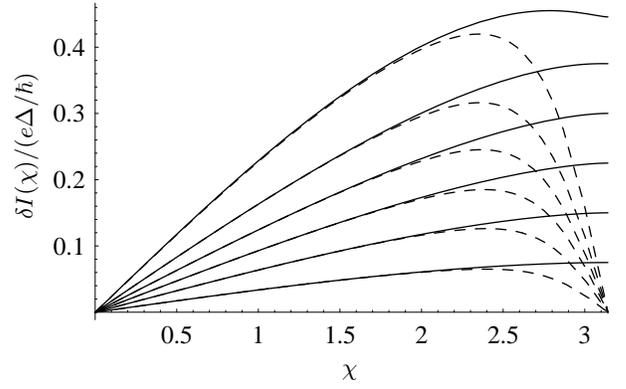}
\caption{Mesoscopic fluctuations of the Josephson current,
$\delta I(\chi)$ vs. phase $\chi$ at zero temperature (solid line)
and at temperature $kT=\minigap/10$ (dashed line). The curves are
plotted for different ratios $\Delta/E_g$: 0 (top), 0.2, 0.5, 1,
2, 5 (bottom). } \label{F:dIT}
\end{figure}

A peculiar feature of the $T=0$ mesoscopic fluctuations
given by Eq.~(\ref{QD-II-cumulant})
is a finite limit at $\chi\rightarrow\pi$:
\be
\label{eq:dIcPI}
   \delta I(\pi)
   =
   \frac{\sqrt{2}e}{\pi\hbar}\frac{\Delta E_g}{\Delta+E_g}
   \sim
   \frac{e E_*}{\hbar},
\ee
while the Josephson current must vanish exactly
at $\chi=\pi$.\cite{Beenakker93,Micklitz}
This is related to the breakdown of the Gaussian treatment
of fluctuations at $\chi\to\pi$.
A more careful analysis of the general nonlinear action
(\ref{eq:action}) should restore the exact relation $I(\pi)=0$,
resulting in vanishing fluctuations for $\chi\to\pi$.\cite{Micklitz}
At a small temperature, $\delta I(\pi)=0$, and the Josephson current
decreases from its typical value (\ref{eq:dIcPI}) to 0
in the small phase range $|\chi-\pi|\lesssim
kT/E_*$, see Fig.~\ref{F:dIT}.

{\em Close to the critical temperature, $T_c$},
when $\Delta(T)\rightarrow 0$, we find
\begin{multline}
  \delta I_c^2
  =
  \left(
    \frac{2ekT}{\hbar}
    \Delta^2(T)E_g^2
  \right)^2
\\ {}
  \times
  \sum_{\epsilon,\epsilon'>0}
  \frac{1}{\epsilon^2\epsilon'^2(\epsilon+\epsilon'+2E_g)^2(\epsilon+E_g)(\epsilon'+E_g)}.
\end{multline}
Thus,
\be
   \delta I_c
   =
   \begin{cases}
     (e \Delta^2(T)/8\hbar k  T_c), & \text{if $kT_c\ll E_g$}, \\
     0.010 e \Delta^2(T) E_g^2/\hbar k^3 T_c^3, & \text{if $E_g\ll kT_c$}.
   \end{cases}
\ee

{\em At intermediate temperatures}\/ such that $E_g\ll kT \ll kT_c$, we find
\begin{equation}
\delta I_c^2
=
\left(\frac{2ekT}{\hbar} E_g^2\right)^2
\sum_{\epsilon,\epsilon'>0}
\frac{1}{\epsilon\epsilon'(\epsilon+\epsilon')^2}.
\end{equation}
Thus, $\delta I_c \simeq 0.12 (e E_g^2 /\hbar k T)$. This result
was also obtained by means of the fourth order perturbation theory in
the tunnel Hamiltonian connecting the wire to the leads, plus random
matrix theory on the statistics of eigenstates in the
lead.\cite{perturbation-theory}

\subsection{Summary of results}
\label{SS:QD-results}

The results of our study are summarized in Table \ref{tab:dot}.
We note that, quite generally, the critical current at low
temperatures is related to the minigap in the dot at $\chi=0$,
$\minigap\sim\min(E_g,\Delta)$, through an Ambegaokar-Baratoff
like formula: $I_c\sim G\minigap/e$, where $G$ is the
normal-state conductance of the system (up to a
logarithmic factor for $E_g\ll\Delta$).

In a normal metallic double-barrier structure, conductance
fluctuations are ``universal": $\delta G=G_Q/2$, while
weak localization corrections vanish at infinitely small mean
level spacing in the dot\cite{iida,beenakker-RMT}.

In an S-QD-S junction,
we find that the WL correction to the Josephson current also vanishes,
while the amplitude of mesoscopic fluctuations can be estimated
as $\delta I_c/I_c\sim \delta G/G$, provided $kT_c\ll E_g$.
In the opposite limit, for poorly conducting barriers, mesoscopic
fluctuations are additionally suppressed.

Note that in the limit $kT\lesssim kT_c\ll E_g$, 
we have $\delta I_c/I_c=\delta G/G$. This should be attributed 
to the fact that in this limit the Josephson relation becomes sinusoidal
and proportional to the conductance of the system. Indeed, using
Eq.~(15) from Ref.~\onlinecite{beenakker91} with $\Delta(T)\ll kT$,
we find $I(\chi)=I_c\sin\chi$, where the critical current is proportional
to the exact conductance before disorder averaging:
$I_c=(e\Delta^2(T)/4\hbar kT) \sum_nT_n = \pi\Delta^2(T)G/4ekT$.


\section{Superconductor-normal metal-superconductor junctions}
\label{sec:S-N-S}

We turn now to the case of a diffusive metallic wire connected to
superconducting leads by transparent interfaces (the interface
resistances are much lower than the resistance of the wire in the
normal state).

Proximity effect in such a geometry can be described by the diffusive
replica $\sigma$-model with the action\cite{Fin,Efetov-book}
\be
\label{S[Q]-wire}
   S[Q]
   = \frac{\pi\nu}{8}
   \int d\mathbf{r}
   \tr \left[ D({\bm \nabla} Q)^2 - 4\epsilon \PauliNambu_3 Q\right],
\ee
where $D$ is the diffusion coefficient.
At the boundaries with superconductors, we require that $Q$ matches
with the $Q_i$ matrices in the leads given by Eq.~(\ref{QS}).

In this Section 
we concentrate on a quasi-one-dimensional geometry, when the length
of the wire, $L$, much exceeds its transverse dimensions.
Then, the spatial dependence of $Q$ is reduced to the dependence
on the coordinate $x$ along the wire, which will be measured in units of $L$.
The action (\ref{S[Q]-wire}) can be written as
\begin{equation}\label{action-wire}
   S[Q]
   =
   \frac{G_N}{16G_Q}
   \int_{-1/2}^{1/2} d x
   \tr \left[ (\nabla Q)^2 - 4\eps \PauliNambu_3 Q\right],
\end{equation}
where $G_N=2\pi G_Q E_T/\delta$ is the normal-state conductance of the wire,
and $\eps=\epsilon/E_T$ stands for Matsubara energies measured in units
of the Thouless energy, $E_T=\hbar D/L^2$.

Following the same line as in Sec.~\ref{sec:S-QD-S}, we derive
the quasiclassical Josephson relation as the saddle point of the
action (\ref{action-wire}) in Sec.~\ref{sec:I-wire}. We find
that the amplitude of the critical current is controlled by the
ratio between $\Delta$ and $E_T$. Then, we express mesoscopic
fluctuations and WL correction to the Josephson current in terms
of Gaussian fluctuations in the vicinity of the non-uniform
saddle point in Sec.~\ref{sec:fluct-wire}.
Results are summarized in Sec.~\ref{S:wire-results}.

\subsection{Josephson relation}
\label{sec:I-wire}

At the saddle point, the matrix $Q_0(x)$ which extremizes the action
(\ref{action-wire}) solves the Usadel equation:
\begin{equation}
   -\nabla(Q_0\nabla Q_0)+\eps [\hat{\tau_3},Q_0]=0.
\end{equation}
With the parametrization (\ref{Q0}), this equation can be reduced to
two differential equations:
\begin{subequations}
\label{usadeltot}
\begin{align}
   \nabla (\sin^2\theta \nabla \phi) = 0 ,
\label{usadel0}
\\
   \nabla^2\theta-2\eps \sin\theta
   - (\nabla \phi)^2\sin\theta\cos\theta = 0 ,
\label{usadel}
\end{align}
\end{subequations}
with the boundary conditions $\theta(\pm 1/2)=\theta_s$
and $\phi(\pm 1/2)=\pm \chi/2$.
The Usadel angles at energies $\pm\eps$ are related by the symmetries:
\be
\label{e-e-symmetries}
  \theta_{-\eps}=\pi-\theta_\eps,
  \qquad
  \phi_{-\eps}=\phi_\eps .
\ee

According to Eq.~(\ref{usadel0}), $J=\sin^2\theta \nabla \phi$
(usually referred to as the spectral current) is constant along the wire.
Integrating then Eq.~(\ref{usadel}), we obtain:
\be
\label{const-motion}
  (\nabla \theta)^2
  =
  4 \eps \left(\cos\theta(0)-\cos\theta\right)
  + \frac{J^2}{\sin^2\theta(0)}-\frac{J^2}{\sin^2\theta} .
\ee
This last equation can be solved in quadratures, the implicit
solution being given in terms of elliptic integrals. Its
form is rather cumbersome and can be found in Appendix \ref{A:eqs}.
In a general case, $J$ and $\theta(0)$ at given boundary conditions
can be found only numerically. Note however that their determination
requires solution of only an algebraic rather than differential equation.

\begin{figure}
\includegraphics{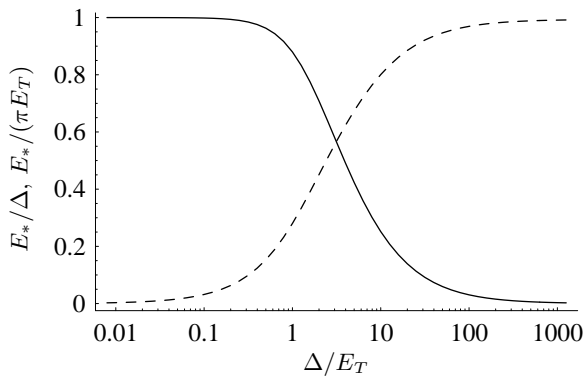}
\caption{The minigap $\minigap$ induced in the normal wire at $\chi=0$
(solid line: in units of $\Delta$,
dashed line: in units of $\pi E_T$)
vs.\ $\Delta/E_T$.}
\label{F:wire-minigap}
\end{figure}

Equation (\ref{const-motion}) can be used to determine the minigap
$\minigap$ induced in the normal region due to the proximity effect.
To this end, one has to perform analytic continuation $\epsilon\to -iE$,
and search for the energy $\minigap$ when $\cos\theta$ first acquires a
finite real part, indicating a nonzero density of states above $\minigap$.
Determination of the minigap simplifies for $\chi=0$. To solve the Usadel
equation we introduce then $\theta=\pi/2+i\psi$ and get the relation
between $E$ and the value $\psi(0)$ at the center of the wire:
\be
\label{minigap}
  \sqrt{\frac{E}{E_T}}
  =
  \int_{\mathop{\rm arctanh}(E/\Delta)}^{\psi(0)}
  \frac{d\psi}{\sqrt{\sinh\psi(0)-\sinh\psi}} .
\ee
Equation (\ref{minigap}) establishes a relation between $E$ and $\psi(0)$.
The value of $\minigap$ should be determined from the condition
that Eq.~(\ref{minigap}) ceases to have real solutions for $\psi(0)$.
For short wires ($\Delta\ll E_T$), $\minigap\approx\Delta$,
whereas for long wires ($\Delta\gg E_T$),
$\minigap\approx 3.122 E_T$,\cite{ZhouCharlat98,OSF01}
in accordance with the general relation $\minigap\sim\min(E_T,\Delta)$.
The dependence of $\minigap$ on $\Delta/E_T$
is shown in Fig.~\ref{F:wire-minigap}.

The action at the saddle point $Q_0(x)$ is given by $nS_0$, where
\be
\label{eq:action-wire-SP}
   S_0
   =
   \frac{G_N}{4G_Q} \sum_\epsilon \int dx
   \left[
     (\nabla \theta)^2+(\nabla \phi)^2 \sin^2\theta-4 \eps \cos\theta
   \right] ,
\ee
Taking the derivative with respect to $\chi$,
integrating by parts, and using the boundary conditions at $x=\pm1/2$
we get $\partial_\chi S_0=(G_N/2G_Q)\sum_\eps J$,
yielding the quasiclassical expression for the supercurrent:
\be
\label{eq:Iav-wire}
   \corr{I(\chi)}_0 = \pi k T\frac{G_N}{e}\sum_\epsilon J.
\ee

Below, we evaluate the current (\ref{eq:Iav-wire}) in different cases.

\subsubsection{Short wire at zero temperature}
\label{SSS:shortT0}

\begin{figure}
\includegraphics[width=70mm]{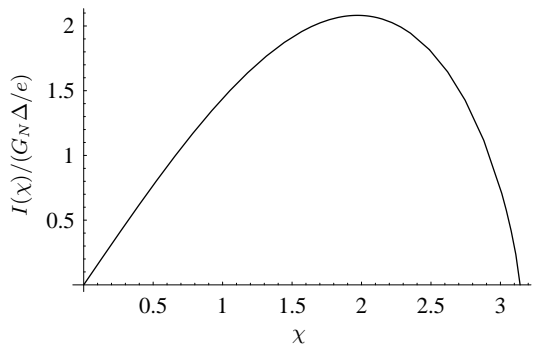}
\caption{Josephson relation $I(\chi)$ (in units of $G_N \Delta/e$)
for a short wire at zero temperature.} \label{F:Ichishort}
\end{figure}

The critical current for a short diffusive wire ($E_T\gg
\Delta$) has been known for a long time.\cite{kulik-omelyanchuk}
In this case, the term proportional to $\eps$ in Eq.~(\ref{usadel})
can be neglected, and the Usadel equation can be solved exactly:
\begin{equation}
   \theta(x)
   =
   \arccos \left[
     \cos \theta(0) \cos \frac{J x}{\sin\theta(0)}
         \right]
  ,
\end{equation}
where
\begin{subequations}
\begin{gather}
\sin\theta(0)=
\frac
     {\sin\theta_s\cos\frac{\chi}{2}}
     {\sqrt{1-\sin^2\theta_s\sin^2\frac{\chi}{2}}},
\\
J=2 \sin\theta(0)
     \arcsin\left(\sin\theta_s\sin\frac{\chi}{2}\right).
\end{gather}
\end{subequations}
Calculating the supercurrent with the help of Eq.~(\ref{eq:Iav-wire})
at $T=0$ we get\cite{kulik-omelyanchuk}
\be
   I(\chi)
   =
   \frac{\pi G_N\Delta}{e}
   \cos\frac{\chi}{2} \,
   \mathop{\rm arctanh} \left( \sin\frac{\chi}{2} \right) .
\ee
The Josephson relation is nonsinusoidal, see Fig.~\ref{F:Ichishort}.
The critical current $I_c=2.082 G_N \Delta/e$ is achieved at the
critical phase $\chi_c=1.255 (\pi/2)$.

\subsubsection{Arbitrary wire at temperature close to $T_c$}
\label{sec:Ic-near-Tc}

At temperatures close to the critical temperature in the
leads, $T_c$, the superconducting gap $\Delta$ in
the leads is small:
$\Delta(T)=[(8\pi^2/7\zeta(3))k^2T_c(T_c-T)]^{1/2}\ll kT$. Thus,
at relevant energies $\epsilon\gtrsim kT\gg \Delta$, the anomalous
(Gor'kov) component of the Green function is small, and the Usadel
equation (\ref{usadel}) can be linearized with respect to
$\sin\theta$. Its solution has the form
\begin{multline}
\label{eq:long-wireT}
  \sin \theta(x) e^{i\phi(x)}
  =
  \frac{\Delta}{|\epsilon|} e^{-i\frac{\chi}{2}}
  \frac{\sinh\kappa(\frac{1}{2}-x)}{\sinh\kappa}
\\ {}
+ \frac{\Delta}{|\epsilon|} e^{i\frac{\chi}{2}}
  \frac{\sinh\kappa(\frac{1}{2}+x)}{\sinh\kappa},
\end{multline}
where $\kappa=\sqrt{2|\eps|}=\sqrt{2|\epsilon|/E_T}$.
Calculating the spectral current with the help of Eqs.~(\ref{theta-Tc})
and (\ref{phi-Tc}), we get
\be
\label{eq:spectralJT}
  J
  =
  \frac{\Delta^2}{\epsilon^2}
  \frac{\kappa}{\sinh\kappa}\sin\chi.
\ee

Close to $T_c$, junctions are classified as short or long
depending on the ratio between $kT_c$ and $E_T$.
For short junctions [$\Delta(T) \ll kT \lesssim kT_c \ll E_T$],
the supercurrent is given by:
\be
   I(\chi)
   =
   \frac{\pi G_N}{4e}\frac{\Delta^2}{kT} \sin\chi .
\ee
For long junctions [$\Delta(T),E_T\ll kT \lesssim kT_c$],
the supercurrent is exponentially suppressed:
\be
   I(\chi)
   \simeq
   \frac{8 G_N}{e}
   \frac{\Delta^2}{\sqrt{2 \pi kT E_T}}
   \exp \left(-\sqrt{\frac{2\pi kT}{E_T}}\right) \sin\chi.
\ee
In both cases the Josephson relation is sinusoidal.

\begin{figure}
\includegraphics{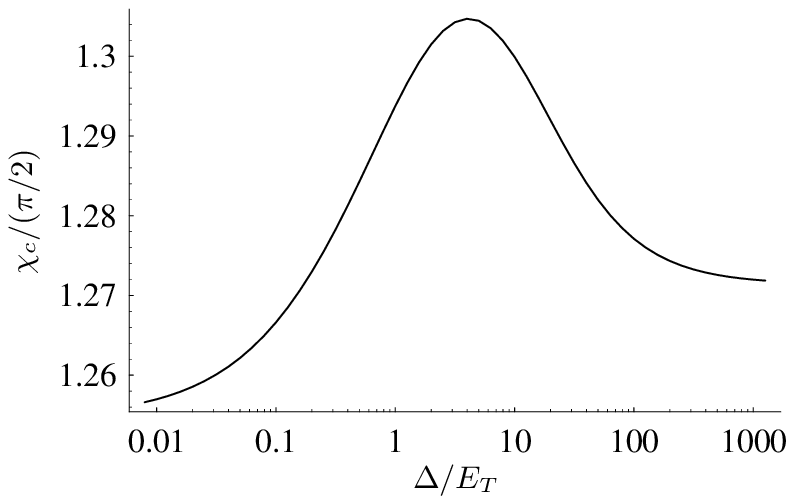}
\hspace*{1mm}\includegraphics{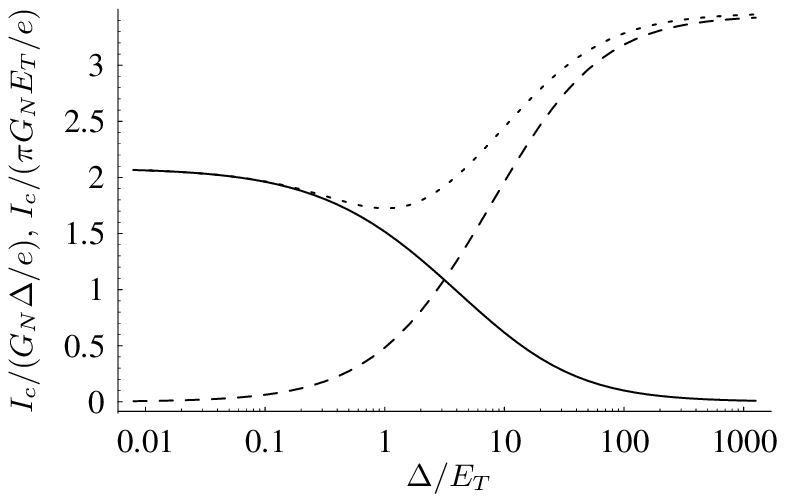}
\caption{(a) The critical phase $\chi_c$ (in units of $\pi/2$) vs.
$\Delta/E_T$ at zero temperature. (b) The critical current $I_c$
(solid line: in units of $G_N\Delta/e$,
dashed line: in units of $\pi G_NE_T/e$,
dotted line: in units of $G_N\minigap/e$)
vs. $\Delta/E_T$ at zero temperature.}
\label{F:Ic-chic-wire}
\end{figure}

\subsubsection{Arbitrary wire at zero temperature}

At $T=0$ the Josephson relation $I(\chi)$ depends only
on the ratio between $\Delta$ and $E_T$.
Short junctions were considered in Sec.~\ref{SSS:shortT0}.
For arbitrary $\Delta/E_T$, the critical current
and critical phase obtained numerically are shown in
Fig.~\ref{F:Ic-chic-wire}.\cite{dubos}

\begin{figure}
\includegraphics{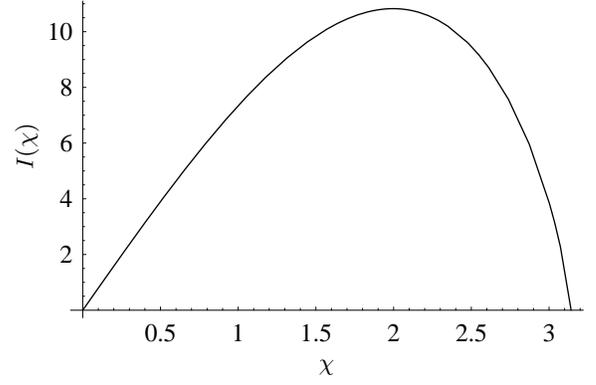}
\caption{Josephson relation $I(\chi)$ (in units of $G_N E_T/e$)
for a long wire at zero temperature.} \label{F:Ichilong}
\end{figure}

Specifically, for long junctions ($E_T\ll\Delta$) the Josephson
relation is still highly nonsinusoidal (see Fig.~\ref{F:Ichilong}),
with the critical current $I_c= 10.83 G_N E_T/e$ achieved
at the critical phase $\chi_c=1.271(\pi/2)$.\cite{dubos}

\subsection{Mesoscopic fluctuations and weak localization correction}
\label{sec:fluct-wire}

Fluctuations near the saddle point $Q_0(x)$ can be parametrized using
Eqs.~(\ref{Q-W}), (\ref{U}) and (\ref{eq:W}). Substituting these
expressions into the action (\ref{action-wire}) and expanding
to the second order in the modes $d(x)$ and $c(x)$, we get similar
to Eq.~(\ref{S2-QD}):
\be
\label{S2-wire}
   S^{(2)}
   =
   \frac{G_N}{4G_Q}
   \int dx
   \sum_{mn}
   \sum_{i=0}^3
   \begin{pmatrix} d^*_{i} & c^*_{i} \end{pmatrix}_{mn}
   \hat{A}_{mn}
   \begin{pmatrix} d_{i} \\ c_{i} \end{pmatrix}_{mn} .
\ee
The form of the operator $\hat A(x)$ is quite cumbersome,
see Eq.~(\ref{A-SNS}) in Appendix \ref{A:A-wire}.
Apart from the second derivative with respect to $x$
it contains also the first derivative. The latter can be
eliminated by a proper unitary transformation [Eq.~(\ref{A-rotation})]
mixing the $d$ and $c$ components of fluctuations.
After such a rotation the matrix $\hat A$ acquires the form\cite{A-rotated}
\begin{multline}
\label{A}
   \hat{A}_{mn}(x)
   =
   - \nabla^2
   + \alpha_{mn}
   + \rho_{mn} \cos(\eta_m+\eta_n) \PauliDC_3
\\
   + \rho_{mn} \sin(\eta_m+\eta_n) \PauliDC_1 ,
\end{multline}
where
\begin{multline}
\label{alpha}
  \alpha_{mn}
  =
  \eps_m\cos\theta_m
     -\frac{1}{4}[(\nabla\theta_m)^2+(\sin\theta_m\nabla\phi_m)^2]
\\ {}
+ \eps_n\cos\theta_n
     -\frac{1}{4}[(\nabla\theta_n)^2+(\sin\theta_n\nabla\phi_n)^2]
     ,
\end{multline}
\vskip -5mm
\begin{multline}
   \rho_{mn}
   =
   \frac12
   \sqrt{(\nabla\theta_m)^2+(\sin\theta_m\nabla\phi_m)^2}
\\ {}
   \times
   \sqrt{(\nabla\theta_n)^2+(\sin\theta_n\nabla\phi_n)^2}
  ,
\end{multline}
and the odd function $\eta_m(x)$ can be obtained by integrating the relation
\be
  \nabla \eta_m
  =
  -
  \frac{2\eps_mJ_m}
    {(\nabla\theta_m)^2+(\sin\theta_m\nabla\phi_m)^2}.
\label{nabla-eta}
\ee

Once the operator $\hat A$ is known, one can use the general
Eqs.~(\ref{eq:IWL}) and (\ref{eq:II}) to calculate the weak
localization correction and mesoscopic fluctuations of the
Josephson current. The only difference compared to the S-QD-S
case is that now $\hat A(x)$ is an operator in the real space,
and the determinant should be calculated with respect to
spacial coordinates as well. In a general situation,
for arbitrary $\Delta/E_T$ and temperatures,
this can be done only numerically, following the procedure
outlined in Appendix \ref{A:eqs}.

Several cases where a simple form of the operator $\hat A(x)$
allows for analytic solution are discussed below.
In Sec.~\ref{SSS:arbT0}, we present the results of numeric
solution for an arbitrary $\Delta/E_T$ at zero temperature.

\begin{figure}
\includegraphics{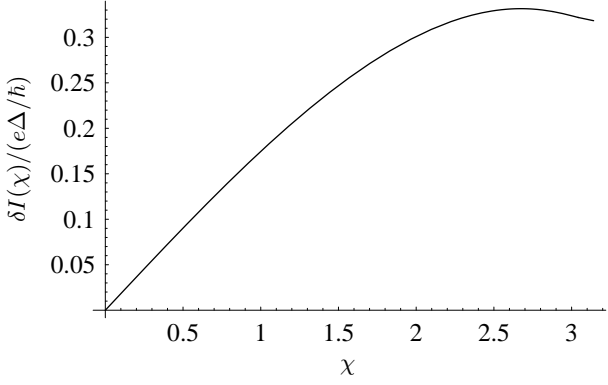}
\caption{Mesoscopic fluctuations $\delta I(\chi)$
(in units of $e\Delta/\hbar$)
for a short wire at $T=0$.}
\label{F:MFshort}
\end{figure}

\subsubsection{Short wire}
\label{SSS:fluct-short}

We start with the simplest case of a short wire, $\Delta\ll E_T$.
This limit was analyzed previously in Refs.~\onlinecite{macedo}
and \onlinecite{beenakker94} using the scattering-matrix approach
to the Josephson current.\cite{beenakker91}

For short wires, the term proportional to $\eps$ in the Usadel
equation can be neglected.
Then, according to Eq.~(\ref{const-motion}),
$(\nabla\theta_m)^2+(\sin\theta_m\nabla\phi_m)^2=4C_m^2$,
where $C_m=J_m/2\sin\theta_m(0)$ is constant in the wire.
With the same accuracy, Eq.~(\ref{nabla-eta}) guaranties that $\eta_m(x)=0$.
As a result, the operator $\hat{A}$ in Eq.~(\ref{A})
becomes diagonal in the $(d,c)$-space, with its diagonal
elements being diffusion operators:
$$
   \hat A_{mn}
   =
   \begin{pmatrix}
   -\nabla^2-(C_m-C_n)^2 & 0 \\
   0 & -\nabla^2-(C_m+C_n)^2
   \end{pmatrix} .
$$
Since fluctuations must vanish in the reservoirs,
the eigenvalues of $-\nabla^2$ are $\pi^2p^2$ ($p=1,2,\dots$),
and the determinant of $A^{\chi_1\chi_2}_{\eps\eps'}$
involved in Eq.~(\ref{eq:II}) can be readily obtained.
As a result, the current-current correlation function
acquires the form
\begin{multline}
\label{II-short}
  \ccorr{I(\chi)I(\chi')}
  =
  -8\left(\frac{e k T}{\hbar}\right)^2
\\ {}
\times
  \frac{\partial^2}{\partial\chi\partial\chi'}
  \sum_{\epsilon\epsilon'}
  \ln \frac{\sin(C-C')}{C-C'} \frac{\sin(C+C')}{C+C'}
  ,
\end{multline}
where $C =
\arcsin[(\Delta/\sqrt{\Delta^2+\epsilon^2})\sin(\chi/2)]$
and $C'$ is given by the same expression with $\eps\to\eps'$
and $\chi\to\chi'$.

At zero temperature, the sums over Matsubara energies reduce
to integrals, and our expression for
$\delta I^2(\chi)=\ccorr{I^2(\chi)}=\var I(\chi)$
becomes equivalent to the result of Refs.~\onlinecite{macedo}
and \onlinecite{beenakker94}.
The equivalence is explicitly demonstrated in Appendix \ref{A:equivalence}.
For small $\chi$, we have an expansion
in powers of $\sin^2(\chi/2)$:
\be
  \delta I^2(\chi)
  =
  \left(\frac{e\Delta}{\hbar}\right)^2
  \frac{\sin^2\chi}{30}
  \left(1+\frac{62}{63}\sin^2\frac{\chi}{2}
  + \dots
  \right) .
\ee
The whole curve $\delta I(\chi)$ at $T=0$
is plotted in Fig.~\ref{F:MFshort}.
Mesoscopic fluctuations of the critical current,
$\delta I_c=\delta I(\chi_c)$, are characterized by
$\delta I_c = 0.30 e\Delta/\hbar$.\cite{macedo,beenakker94}

At zero temperature, the magnitude of mesoscopic fluctuations
remains finite at $\chi=\pi$, $\delta I(\pi) = e\Delta/\pi\hbar$,
contradicting the general symmetry requirement
of vanishing $I(\chi)$.\cite{Beenakker93,Micklitz}
This is related to the breakdown of the Gaussian treatment
of fluctuations at $\chi\to\pi$.
A more careful analysis of the general nonlinear action
(\ref{action-wire}) should restore the exact relation $I(\pi)=0$,
resulting in vanishing fluctuations for $\chi\to\pi$.\cite{Micklitz}
The region
where the exact treatment should modify the result obtained in the
Gaussian approximation can be estimated as $|\chi-\pi|\lesssim G_Q/G_N \ll1$.
Finite temperatures render $\delta I(\pi)=0$, and shrink the region
of strong non-Gaussian fluctuations near $\chi=\pi$,
which disappears at $kT\gg(G_Q/G_N)|\chi-\pi|\Delta$.

\begin{figure}
\includegraphics{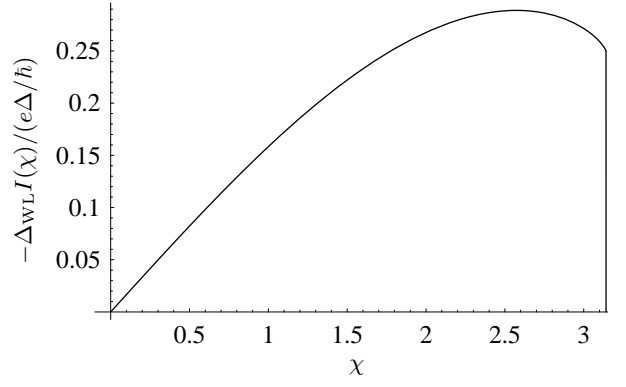}
\caption{Weak localization correction $-\Delta_\text{WL}I(\chi)$
(in units of $e\Delta/\hbar$) for a short wire at $T=0$.}
\label{F:WLshort}
\end{figure}

The weak localization correction can be evaluated
with the help of Eq.~(\ref{eq:IWL}). Here, only $c$-modes
contribute to the result, and
\be
\label{eq:short-WL}
  \Delta_{\text{WL}} I(\chi)
  =
  \frac{e k T}{\hbar}
  \frac{\partial}{\partial\chi}
  \sum_{\epsilon}
  \ln \frac{\sin 2C}{2C} .
\ee
At zero temperature, we obtain:
\be
\Delta_\text{WL} I(\chi)
=
-\frac{e \Delta}{6\hbar} \sin\chi
\left( 1+\frac{7}{15}\sin^2\frac{\chi}{2}+\dots \right).
\ee
in agreement with Eq.~(3.33) of Ref.~\onlinecite{macedo}.
The whole dependence $\Delta_\text{WL}I(\chi)$ at $T=0$
is shown in Fig.~\ref{F:WLshort}.
The WL correction to the critical current is
$\Delta_\text{WL}I_c = -0.266 e\Delta/\hbar$.

At zero temperature, the WL correction (\ref{eq:short-WL}) is discontinuous
at $\chi=\pi$: $\Delta_\text{WL}I(\pi\mp0)=\mp e\Delta/4\hbar$,
which is an artefact of the Gaussian approximation employed.
The situation here is completely analogous to the situation with
mesoscopic fluctuations discussed above.

\subsubsection{Arbitrary wire at temperatures close to $T_c$}
\label{mesofluctT}

In Sec.~\ref{sec:Ic-near-Tc} we have seen that calculation of the
quasiclassical Josephson current simplifies at temperatures close
to $T_c$. The same occurs for mesoscopic fluctuations and WL correction.
In order to calculate them,
we decompose the operator $\hat{A}=\hat{A}_0+\hat{V}$
in Eq.~(\ref{A}) into a sum of the diffusion operator in the normal state,
$\hat{A}_0=-\nabla^2+|\eps_1|+|\eps_2|$, and the perturbation
$\hat V = {\cal O}(\Delta^2)$.
Expanding $\tr\ln\hat{A}$ in powers of $\hat{V}$ we
rewrite Eqs.~(\ref{eq:IWL}) and (\ref{eq:II}) as
\begin{gather}
\label{TWL}
   \Delta_{\text{WL}} I(\chi)
   =
   - \frac{e k T}{\hbar}
   \frac{\partial}{\partial\chi}
   \sum_{\epsilon} \Tr G \PauliDC_3 V_{\epsilon,-\epsilon}^{\chi\chi} ,
\\
  \ccorr{I(\chi_1) I(\chi_2)}
  =
  4\left(\frac{e k T}{\hbar}\right)^2
  \frac{\partial^2}{\partial\chi_1\partial\chi_2}
  \sum_{\epsilon_1\epsilon_2} \Tr (G \hat V)^2 ,
\label{TII}
\end{gather}
where $\hat G=\hat A_0^{-1}$ is the Green function of the diffusion operator:
$[-\nabla^2+|\eps_1|+|\eps_2|]G_{\eps_1\eps_2}(x,y)=\delta(x-y)$,
and $\Tr$ implies tracing over coordinates as well.
Explicitly,
\be
\label{G}
  G_{\eps_1\eps_2}(x,y)
  =
  \frac{\sinh\lambda(\frac{1}{2}+m) \sinh\lambda(\frac{1}{2}-M)}
       {\lambda \sinh \lambda},
\ee
where $m=\min(x,y)$, $M=\max(x,y)$, and $\lambda=\sqrt{|\eps_1|+|\eps_2|}$.

We start with the {\em weak localization correction}.
Expanding
\be
\label{V}
  \hat{V}(x)=V^0(x)+V^1(x)\PauliDC_1+V^3(x)\PauliDC_3,
\ee
with $V^i(x)$ given by Eqs.~(\ref{V0})--(\ref{V3}),
we get for the expression entering Eq.~(\ref{TWL}):
\be
  \Tr G \PauliDC_3 V_{\epsilon,-\epsilon}^{\chi\chi}
  =
  2 \int_{-1/2}^{1/2} dx \, G_{\eps,-\eps}(x,x) (V^3(x))_{\epsilon,-\epsilon}^{\chi\chi} .
\ee
Evaluating the integral using Eqs.~(\ref{G}) and (\ref{V3}) we obtain
the weak localization correction:
\be
  \Delta_{\text{WL}} I(\chi)
  =
  - \frac{ekT}{8\hbar} \sin\chi
  \sum_{\epsilon} \frac{\Delta^2}{\epsilon^2}
  \frac{\cosh2\kappa}{\sinh^3\kappa}
  (2\kappa-\tanh2\kappa) ,
\ee
where $\kappa=\sqrt{2|\eps|}=\sqrt{2|\epsilon|/E_T}$.

For short junctions ($kT_c\ll E_T$),
\be
  \Delta_{\text{WL}} I(\chi)
  =
  - \frac{e\Delta^2}{12\hbar kT} \sin\chi ,
\ee
and
$\Delta_{\text{WL}}I(\chi)/I(\chi) = - G_Q/3G_N$.

For long junctions ($E_T\ll kT_c$),
\be
  \Delta_{\text{WL}} I(\chi)
  =
  - \frac{4}{\pi}
  \frac{e\Delta^2}{\hbar\sqrt{2\pi kT E_T}} e^{-\sqrt{2\pi kT/E_T}}
  \sin\chi ,
\ee
and
$\Delta_{\text{WL}}I(\chi)/I(\chi) = - G_Q/2G_N$.

\begin{figure}
\includegraphics{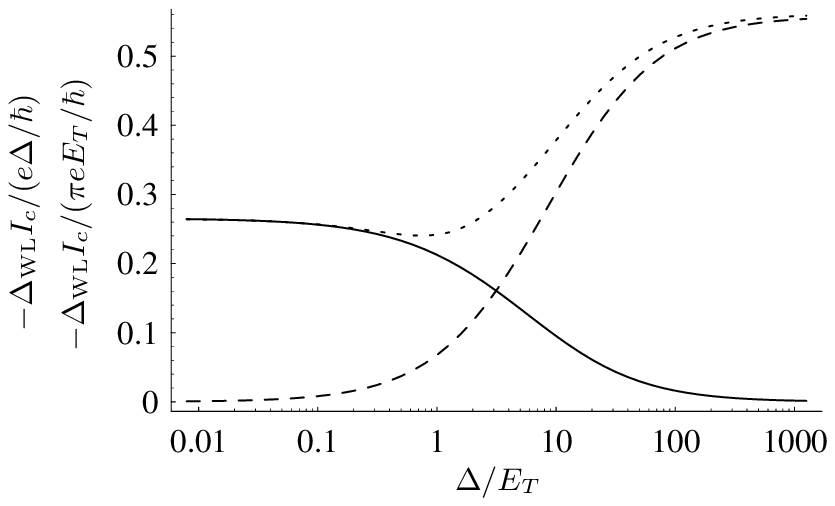}
\hspace*{3.0mm}\includegraphics{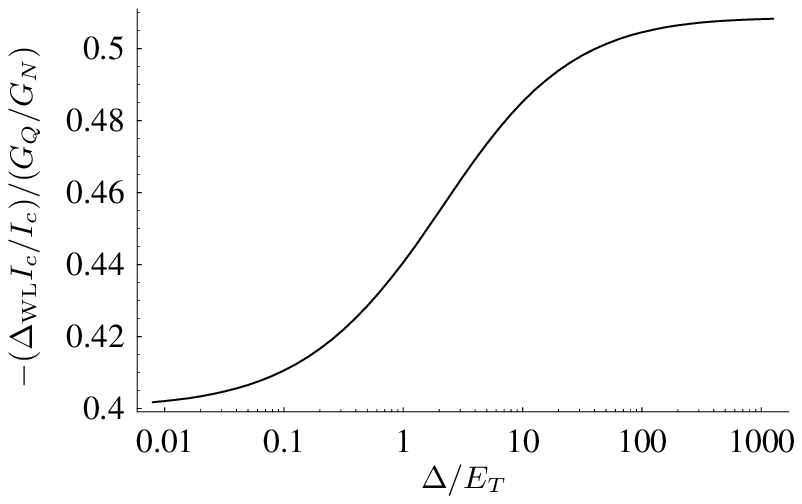}
\caption{(a) Weak localization correction to the critical current
$-\Delta_{\text{WL}} I_c$
(solid line: in units of $e\Delta/\hbar$,
dashed line: in units of $\pi e E_g/\hbar$,
dotted line: in units of $e\minigap/\hbar$)
vs.\ $\Delta/E_T$ at zero temperature.
(b) The ratio $-\Delta_{\text{WL}} I_c/I_c$ (in units of $G_Q/G_N$)
vs.\ $\Delta/E_T$ at zero temperature.}
\label{F:WL}
\end{figure}

{\em Mesoscopic fluctuations}\/ are calculated in Appendix~\ref{A:highT}
with the help of Eq.~(\ref{TII}). The result has the form
\begin{multline}
\label{eq:mesoT}
  \ccorr{I(\chi_1) I(\chi_2)}
  =
  4\left(\frac{e k T}{\hbar}\right)^2
  \sin\chi_1\sin\chi_2
\\ {}
  \times
  \sum_{\epsilon_1\epsilon_2}
  \frac{\Delta^4}{\epsilon_1^2\epsilon_2^2}
  \left( \frac{\kappa_1}{\sinh\kappa_1} \right)^2
  \left( \frac{\kappa_2}{\sinh\kappa_2} \right)^2
  \Upsilon_{\eps_1\eps_2} ,
\end{multline}
where the function $\Upsilon_{\eps_1\eps_2}$
is defined in Eq.~(\ref{upsilon}).

For short junctions ($kT_c\ll E_T$),
one can take the limit $\kappa_1,\kappa_2,\lambda\to 0$
and get $\Upsilon_{\eps_1\eps_2}=1/30$, leading to
\be
  \ccorr{I(\chi_1) I(\chi_2)}
  =
  \frac{1}{120}
  \left(\frac{e\Delta^2}{\hbar k T}\right)^2
  \sin\chi_1\sin\chi_2 .
\ee
This result could have been deduced already from the exact
result (\ref{II-short}) for short junctions.
The relative fluctuations of the Josephson current are
$\delta I(\chi)/I(\chi) = \sqrt{2/15}\,G_Q/G_N$.

For long junctions ($E_T\ll kT_c$),
the sums in Eq.~(\ref{eq:mesoT}) are dominated
by the lowest Matsubara frequencies
$\epsilon_1,\epsilon_2=\pm\pi T$, and
$\Upsilon_{\pi T,\pi T}=e^{2\kappa}/128\kappa^2$
with $\kappa=\kappa_1=\kappa_2$.
Hence
\be
  \ccorr{I(\chi_1) I(\chi_2)}
  =
  \frac{4e^2\Delta^4}{\pi^3\hbar^2 k T E_T}
  e^{-\sqrt{8\pi kT/E_T}}
  \sin\chi_1\sin\chi_2 ,
\ee
and the relative mesoscopic fluctuations are
$\delta I(\chi)/I(\chi) = G_Q/2\sqrt{2}G_N$.

\begin{figure}
\hspace*{0.5mm}\includegraphics{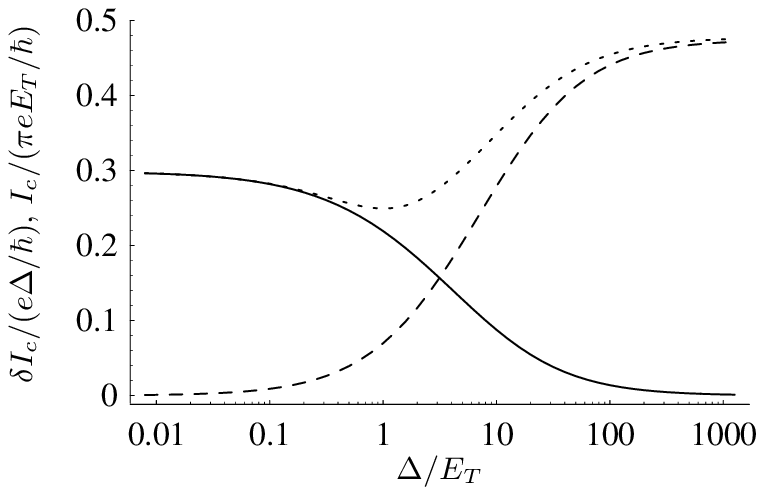}
\vskip2mm
\includegraphics{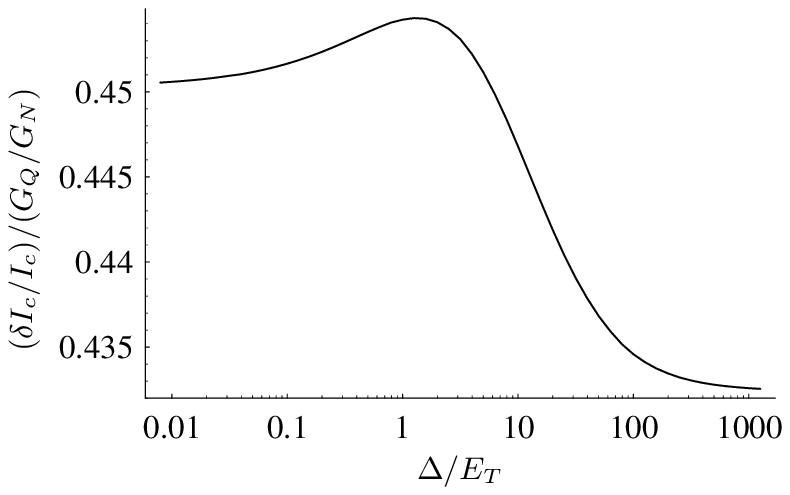}
\caption{(a) The amplitude of mesoscopic fluctuations of the critical current
$\delta I_c$
(solid line: in units of $e\Delta/\hbar$,
dashed line: in units of $\pi eE_T/\hbar$,
dotted line: in units of $e\minigap/\hbar$)
vs.\ $\Delta/E_T$ at zero temperature.
(b) The ratio $\delta I_c/I_c$ (in units of $G_Q/G_N$)
vs.\ $\Delta/E_T$ at zero temperature.}
\label{F:MF}
\end{figure}

\subsubsection{Arbitrary wire at zero temperatures}
\label{SSS:arbT0}

Finally, we discuss the case of an arbitrary wire
at zero temperature, when the problem can be solved only numerically.
Here, the amplitude of mesoscopic fluctuations, $\delta I(\chi)$,
and WL correction, $\Delta_\text{WL}I(\chi)$, are functions of two
parameters: the ratio $\Delta/E_T$ and the phase difference $\chi$.

The dependencies of the WL correction to the critical current
and its mesoscopic fluctuations on $\Delta/E_T$
are presented in Figs.~\ref{F:WL} and \ref{F:MF}, respectively.
One can clearly see the crossover from the short to long limits
at $\Delta\sim5E_T$. Note that the relative magnitude of mesoscopic
fluctuations shown in Fig.~\ref{F:MF}(b) is nearly insensitive
to the wire's length and can be approximated by $\delta I_c/I_c\sim0.44G_Q/G_N$.

\begin{figure}
\includegraphics{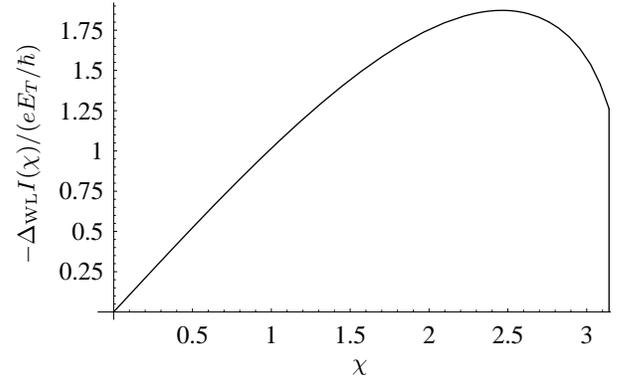}
\caption{Weak localization correction $-\Delta_{\text{WL}} I(\chi)$
(in units of $eE_T/\hbar$)
vs. $\chi$ for long wires at zero temperature.}
\label{F:WL-long}
\end{figure}

\begin{figure}
\includegraphics{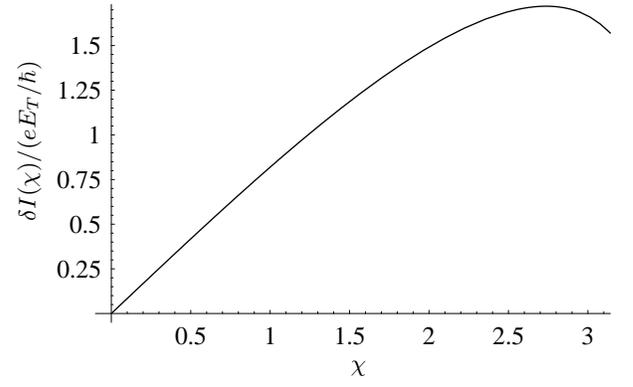}
\caption{Mesoscopic fluctuations $\delta I(\chi)$
(in units of $eE_T/\hbar$) vs.\ $\chi$ for long wires at zero temperature.}
\label{F:MF-long}
\end{figure}

{\em For long junctions}\/ ($E_T\ll\Delta$),
the energy scale for mesoscopic fluctuations
is set by the Thouless energy: $\delta I_c = 1.490 e E_T/\hbar$
and $\Delta_\text{WL} I_c = -1.754 e E_T/\hbar$.
The $\chi$-dependences of the WL correction and mesoscopic fluctuations
in this regime are shown in Figs.~\ref{F:WL-long} and \ref{F:MF-long},
respectively.
These plots are inaccurate in the vicinity of $\chi=\pi$,
cf.\ discussion in Sec.~\ref{SSS:fluct-short}.

Note that our result for the magnitude of mesoscopic fluctuations
of the critical current for long wires is 2.5 times
larger than the prediction of Ref.~\onlinecite{altshuler87}.

\begin{table*}
\caption{\label{tab:wire}
Summary of results for a wire with conductance $G_N$ contacted
to superconducting leads.}
\begin{ruledtabular}
\begin{tabular}{ccccccc}
& $\chi_c/(\pi/2)$ & $e I_c/G_N$ & $\hbar\delta I_c/e$ & $g_N\delta I_c/I_c$ & $\hbar\Delta_\text{WL} I_c/e$ & $g_N\Delta_\text{WL} I_c/I_c$ \\
\hline
$\Delta\ll E_T,\, T=0$ & $1.255$ & $2.082\Delta{}^{[\text{\onlinecite{kulik-omelyanchuk}}]}$ & $0.30\Delta{}^{[\text{\onlinecite{macedo,beenakker94}}]}$ & 0.45 & $-0.266\Delta{}^{[\text{\onlinecite{macedo}}]}$ & $-0.401$ \\
$E_T\ll \Delta,\, T=0$ & $1.271$ & $10.83 E_T{}^{[\text{\onlinecite{dubos}}]}$ & $1.490 E_T$ & 0.432 & $-1.754E_T$ & $-0.509$ \\
arb.\ $E_T/\Delta$, $T=0$ & Fig.\ref{F:Ic-chic-wire}(a) & Fig.\ref{F:Ic-chic-wire}(b)${}^{[\text{\onlinecite{dubos}}]}$ & Fig.~\ref{F:MF}(a) & Fig.~\ref{F:MF}(b) & Fig.~\ref{F:WL}(b) & Fig.~\ref{F:WL}(b) \\
$kT \lesssim kT_c \ll E_T$ & $1$ & $\pi \Delta^2(T)/4kT_c{}^{[\text{\onlinecite{kulik-omelyanchuk}}]}$ & $\Delta^2(T)/\sqrt{120}kT$ & $\sqrt{2/15}$ & $-\Delta^2(T)/12kT$ & $-1/3$ \\
$E_T\ll kT \lesssim kT_c$ & $1$ & $8\Delta^2(T)/{\cal E}{}^{[\text{\onlinecite{ZaikinZharkov81}}]}$ & $2^{3/2}\Delta^2(T)/{\pi\cal E}$ & $1/(2\sqrt{2})$ & $-4\Delta^2(T)/\pi{\cal E}$ & $-1/2$ \\
\end{tabular}
\end{ruledtabular}
\vskip 3pt
\footnotesize
Here, $g_N=G_N/G_Q$, $\Delta(T)=[(8\pi^2/7\zeta(3))k^2T_c(T_c-T)]^{1/2}$,
and ${\cal E}=\sqrt{2\pi kTE_T} \exp(\sqrt{2\pi kT/E_T})$.
\hfill{}
\end{table*}

\subsection{Summary of results}
\label{S:wire-results}

The results of this Section are summarized in Table \ref{tab:wire}.
The critical current at low temperatures
is related to the minigap in the normal region at $\chi=0$,
$\minigap\sim\min(E_T,\Delta)$, through an
Ambegaokar-Baratoff like formula: $I_c\sim G_N\minigap/e$.
We see that, quite generally, mesoscopic fluctuations and
WL correction to the Josephson current are suppressed by the
factor $G_Q/G_N$:
$\delta I_c \sim - \Delta_\text{WL}I_c \sim e\minigap/\hbar$,
where the precise coefficients in this expression depend
on the wire's length and temperature.
Note that contrary to the S-QD-S cases considered
in Sec.~\ref{sec:S-QD-S}, these coefficient are always of
the order of 1.
This should be compared to the behavior of the conductance.
In a normal metallic wire, conductance fluctuations and its weak localization
correction are also suppressed by the same factor $G_Q/G_N$,
but the coefficients are ``universal'':
$\delta G_N=\sqrt{2/15}\,G_Q$
and $\Delta_\text{WL}G_N=-G_Q/3$.

Note that for short wires at $T$ close to $T_c$,
the relative WL correction and mesoscopic fluctuations of $I_c$
coincide with those of the conductance. This is due to the relation
$I(\chi)\propto G\sin\chi$, see discussion in the end of 
Sec.~\ref{SS:QD-results}.

\section{Mesoscopic fluctuations in 2D and 3D geometries}
\label{S:23D}

\subsection{Mesoscopic fluctuations of the Josephson current through
a 2D electron gas}
\label{SS:2D}

In Ref.~\onlinecite{takayanagi}, mesoscopic fluctuations of the critical current
had been measured in junctions formed by a two-dimensional electron gas whose
width $W$ was much longer than the distance $L$ between the superconducting
electrodes. In this case, the results of Sec.~\ref{sec:S-N-S} cannot be applied
since transverse diffusive modes become relevant. In this Section we take them
into account and calculate rms $\delta I_c$ for the two-dimensional geometry.

Mesoscopic fluctuations can be obtained with the help of the general
formula (\ref{eq:II}), where now the operator $\hat A$ reads
\be
\label{A2D}
  \hat A^\text{(2D)}(x,y)
  =
  - \frac{\partial^2}{\partial y^2}
  + \hat A(x) .
\ee
Here, $\hat A(x)$ stands for the operator (\ref{A}) for the quasi-1D wire,
and $-1/2<x<1/2$ ($0<y<W/L$) is the dimensionless coordinate along
(perpendicular to) the junction.

The eigenvalues of the operator (\ref{A2D}) are given by
\be
  \lambda_{m,i}
  =
  \frac{\pi^2 m^2}{\eta^2} + \lambda_i ,
\ee
where $\eta=W/L$, and $\lambda_i$ are the eigenvalues
of the operator $\hat A(x)$ (we suppressed the indices $\chi_1,\chi_2$
and $\eps,\eps'$ for brevity).
For wires, only the zeroth transverse diffusive mode with $m=0$
was relevant, whereas for the film geometry one has to sum over all $m$'s.
As a result, we can express the current cumulant in terms of the
spectrum of the operator $\hat A(x)$:
\be
\label{eq:II-2D}
\ccorr{I(\chi_1) I(\chi_2)}
=
-8\left(\frac{e k T}{\hbar}\right)^2
\frac{\partial^2}{\partial\chi_1\partial\chi_2}
\sum_{\epsilon\epsilon'} \tr F( A^{\chi_1\chi_2}_{\epsilon\epsilon'} ) ,
\ee
where
\be
\label{F-2D}
  F(\lambda)
  =
  \ln \biggl( \frac{\sqrt{\lambda}}{\eta} \sinh \eta\sqrt{\lambda} \biggr)
\ee
is a generalization of the function $\ln \lambda$ relevant for wires
to arbitrary ratios $\eta=W/L$.

We analyze Eqs.~(\ref{eq:II-2D}) and (\ref{F-2D}) in the experimentally
relevant limit\cite{takayanagi} of wide ($W\gg L$) and long
($E_T\equiv\hbar D/L^2\ll\Delta$) junctions, at small temperatures ($kT\ll E_T$).
Then $F(\lambda)\approx\eta\sqrt\lambda$,
and the spectral sum should be calculated numerically, analogously
to the quasi-one-dimensional situation, see Sec.~\ref{SSS:arbT0}.
We obtain for the rms of the critical current fluctuations
(at $\chi_c=1.27 \pi/2$):
\be
\label{fluct-2D}
  \delta I_c = 1.5 \frac{eE_T}{\hbar} \sqrt{\frac{W}{L}} .
\ee

A similar equation with the prefactor 2.2
was used in Ref.~\onlinecite{takayanagi} [see Eq.~(2) there]
as the theoretical estimate for the critical current fluctuations.
Though the two-dimensional case was not considered by Altshuler
and Spivak\cite{altshuler87}, it was claimed
that Eq.~(2) of Ref.~\onlinecite{takayanagi}
can be obtained from the three-dimensional result of
Ref.~\onlinecite{altshuler87}.
We could not follow the derivation of Eq.~(2) in Ref.~\onlinecite{takayanagi}
but would like to emphasize that even the three-dimensional result
for $\delta I_c$ from Ref.~\onlinecite{altshuler87} is overestimated in Eq.~(1)
of Ref.~\onlinecite{takayanagi} by the factor of $\pi^2$ due a different
definition of the Thouless energy. So we expect that the 2D result for
$\delta I_c$ obtained within the Altshuler-Spivak approach would have the
form (\ref{fluct-2D}) with the prefactor smaller than 1.

Thus, even though the approach of Ref.~\onlinecite{altshuler87} generally
underestimates the magnitude of critical current fluctuations, the
theoretical prediction of Ref.~\onlinecite{takayanagi} overestimated them
by the factor of 1.5.

With the help of Eq.~(\ref{fluct-2D}) we have a better explanation
of experimental results (in the weakly localized regime I)
from Ref.~\onlinecite{takayanagi}:
$\delta I_c^\text{theor}=45$ nA, $\delta I_c^\text{exper}=15$ nA.
Still, discrepancy by the factor of 3 remains.

For completeness, we present also the result for wide ($W\gg L$)
and short ($\Delta\ll E_T$) junctions at $T=0$. It can be obtained
using the spectrum found in Sec.~\ref{SSS:fluct-short}. We get
for the rms of the critical current fluctuations:
\be
  \delta I_c = 0.26 \frac{e\Delta}{\hbar} \sqrt{\frac{W}{L}}.
\ee

\subsection{3D case}
\label{SS:3D}

Finally, we mention that the above results for the 2D case can be easily
generalized to the 3D case.
To be specific, we consider the limit of wide junctions,
when both transverse dimensions are large: $W_y,W_z\gg L$.
Then, mesoscopic fluctuations can be found with the help
of Eq.~(\ref{eq:II-2D}) with
$F(\lambda)=(W_yW_z/4\pi L^2)\lambda\ln M/\lambda$,
where $M\sim L/l$ is the high-momentum cutoff
[it drops from the answer since
$\partial^2\tr\hat A/\partial\chi\partial\chi'=0$,
see Eq.~(\ref{A})].

For long junctions ($E_T\ll\Delta$) at $T=0$, numerical integration
leads to Eq.~(\ref{II-3D-long}).
Note that the result of Ref.~\onlinecite{altshuler87} obtained in the same
limit contains the numerical factor $\sqrt{15\zeta(5)/\pi^3}=0.71$
instead of 2.0.
Again, we see that the approach of Ref.~\onlinecite{altshuler87}
underestimates the magnitude of mesoscopic fluctuations of $I_c$
by a factor of 2.8.

For short junctions ($\Delta\ll E_T$) at $T=0$, we obtain
\be
  \delta I_c = 0.28 \frac{e\Delta}{\hbar} \sqrt{\frac{W_yW_z}{L^2}}.
\ee

\section{Conclusion}

In this work, we used the replica $\sigma$-model technique to
describe mesoscopic fluctuations and weak localization correction
to the equilibrium supercurrent in Josephson junctions
formed of a metallic wire between superconducting leads.
We considered two types of junctions: a chaotic dot coupled to
superconductors by tunnel barriers (S-QD-S)
and a diffusive wire (SNS) with transparent NS interfaces.
In both cases we calculated the amplitude of supercurrent fluctuations
and the weak localization correction to the average current $I(\chi)$
in different temperature regions at
arbitrary ratios between $\Delta$
and $E_\text{dwell}$ (given by $E_g$ for an S-QD-S junction,
and by $E_T$ for an SNS junction).

For a quasi-one-dimensional SNS junction,
we have found that mesoscopic corrections to the
quasiclassical Josephson current are ``nearly universal'':
$\delta I_c/I_c \sim - \Delta_\text{WL}I_c/I_c \sim G_Q/G$,
where the exact coefficients in these relations are of the order
of 1, being slightly dependent on the parameters of the junction.
For a double-barrier S-QD-S junction, the weak localization correction
vanishes, while mesoscopic fluctuations are ``less universal'':
$\delta I_c/I_c \sim G_Q/G$ for junctions with $kT_c\ll E_g$,
and is additionally suppressed for junctions with $E_g\ll kT_c$.

We also demonstrate that the approach of Ref.~\onlinecite{altshuler87}
systematically underestimates the magnitude of mesoscopic fluctuations
of the critical current by factors around 2.5--2.8, both in the
quasi-one-dimensional and three-dimensional cases.

Theoretical predictions for the mesoscopic fluctuations of the
critical current should be compared with experimental
results.\cite{takayanagi,defranceschi}
In Ref.~\onlinecite{takayanagi}, mesoscopic fluctuations of $I_c$
were studied in a geometry of wide ($W\gg L$) long ($E_T\ll\Delta$)
bar of a two-dimensional electron gas. Experimentally observed
fluctuation magnitude (in the weakly localized regime I)
is three times smaller than our result (\ref{fluct-2D}).
Experiment of Ref.~\onlinecite{defranceschi} refers to the
quasi-one-dimensional short ($\Delta\ll E_T$) wires,
and should be compared with the result\cite{macedo,beenakker94}
$\delta I_c=0.30e\Delta/\hbar$. Again, theoretical prediction ($\sim 7$ nA)
appears to be several times larger than the experimentally measured
magnitude.
This systematic discrepancy might be attributed
either to the effect of electron-electron
interaction or to non-ideal transparencies of the NS interfaces.

Among the motivations to study mesoscopic fluctuations, it was
suggested that they may induce the sign change of the critical
current ($\pi$-junction behavior). \cite{spivak-kivelson} However,
such possibility was discarded in Ref.~\onlinecite{titov} if
time-reversal symmetry is preserved in the normal part, and if
interactions are absent. If the normal part is ferromagnetic,
time-reversal symmetry is broken. Then, with strong barriers
between the wire and leads, it was shown that mesoscopic
fluctuations are dominant in the Josephson relation compared to
the quasiclassical contribution.\cite{zyuzin} Within our approach
this study could be easily reconsidered and extended  by adding an
exchange energy term in the Hamiltonian of the wire. A strong
Coulomb blockade in the central dot was also shown to provide a
mechanism of $\pi$-junction formation via
fluctuations.\cite{guinea}
We note that the replica $\sigma$-model
is also appropriate for taking electron-electron interactions into account.
This could be the subject of future study.

\acknowledgments

We thank M.~V.~Feigelman, Ya.~V. Fominov, and L.~I.~Glazman
for useful discussions.
The work of M.A.S. was supported by the RFBR under grant No.\ 04-02-16998,
and the Russian Science Support Foundation.


\appendix

\section{Solution of the Usadel equations for a wire}
\label{A:eqs}

\subsection{Determination of the spectral current $J$ and $\theta(0)$}

For a fixed $\eps>0$, $\theta_s$ and $\chi$,
the values of the spectral current $J$
and the Usadel angle $\theta_0\equiv\theta(0)$ in the middle of the wire
are determined from two equations:
\begin{subequations}
\begin{gather}
   \sqrt{\eps}
   =
   \int_{\theta_0}^{\theta_s}
   \frac{d\theta}
    {{\cal R}(\theta)},
\\
   \chi
   =
   2 \sqrt{\alpha}
   \int_{\theta_0}^{\theta_s}
   \frac{1}{\sin^2\theta}
   \frac{d\theta}
    {{\cal R}(\theta)},
\end{gather}
\end{subequations}
where ${\cal R}(\theta)=[\cos\theta_0-\cos\theta
+ \alpha(\sin^{-2}\theta_0-\sin^{-2}\theta)]^{1/2}$ and
$\alpha=J^2/4\eps$.
These integrals can be converted to the standard elliptic integrals:
\begin{subequations}
\label{ell-eq-12}
\be
\label{ell-eq-1}
   \sqrt{\eps}
   =
   \frac{2 F(\varphi,k)}{\sqrt{a-c}} ,
\ee
\begin{multline}
\label{ell-eq-2}
   \chi
   =
   \frac{4 \sqrt\alpha F(\varphi,k)}{(1-a^2)\sqrt{a-c}}
   +
   \frac{2 \sqrt\alpha (b-a)}{\sqrt{a-c}}
\\
   \quad\times
   \left[
     \frac{\Pi(\varphi,k^2\frac{1-a}{1-b},k)}{(1-a)(1-b)}
   - \frac{\Pi(\varphi,k^2\frac{1+a}{1+b},k)}{(1+a)(1+b)}
   \right] ,
\end{multline}
\end{subequations}
where
\begin{subequations}
\begin{gather}
   \biggl\{ \begin{matrix}a\\c\end{matrix} \biggr\}
   = \frac{\alpha}{2\sin^2\theta_0}
   \pm \sqrt{1+\frac{\alpha\cos\theta_0}{\sin^2\theta_0}+\frac{\alpha^2}{4\sin^4\theta_0}}
   ,
\\
   b = \cos\theta_0,
   \qquad
   k
   =
   \sqrt{\frac{b-c}{a-c}},
\\
   \varphi
   =
   \arcsin\sqrt{\frac{(a-c)(b-\cos\theta_s)}{(b-c)(a-\cos\theta_s)}}
.
\end{gather}
\end{subequations}
Note that our definition of elliptic integrals $F(\varphi,k)$
and $\Pi(\varphi,n,k)$ coincides with that of Ref.~\onlinecite{GradsteinRyzhik}.
The same functions are often defined in a different way\cite{AbramowitzStegun}:
in Mathematica, e.g., $F(\varphi,k)=\text{\tt EllipticF}[\varphi,k^2]$,
and $\Pi(\varphi,n,k)=\text{\tt EllipticPi}[n,\varphi,k^2]$.

Equations (\ref{ell-eq-12}) should be solved
numerically to obtain $\alpha$ and $b$ (and hence $J$ and $\theta_0$)
for given $\eps>0$, $\theta_s$ and $\chi$.
The Usadel angles for $\eps<0$ can be obtained
from Eqs.~(\ref{e-e-symmetries}).

\subsection{Determination of $\theta(x)$ and $\eta(x)$}

For each $\eps$ we first have to find
$\theta_0$ and $\alpha$ as described above.
Then for a spacial point $x_i$, the value $\theta_i=\theta(x_i)$
can be found from the equation
\be
   |x_i|
   =
   \frac{F(\varphi_i,k)}{\sqrt{\eps}\sqrt{a-c}} ,
\qquad
   \varphi_i
   =
   \arcsin\sqrt{\frac{(a-c)(b-\cos\theta_i)}{(b-c)(a-\cos\theta_i)}} ,
\label{xi}
\ee
which can be solved as
\be
   \cos\theta_i = \frac{b-aY}{1-Y},
\qquad
   Y = \frac{b-c}{a-c} \sn\nolimits^2(|x_i|\sqrt{\eps(a-c)},k) ,
\ee
where $\sn(u,k)$ is the Jacobi elliptic function\cite{GradsteinRyzhik}.
In Mathematica, e.g., $\sn(u,k)=\text{\tt JacobiSN}[u,k^2]$.

Then we determine $\eta_i=\eta(x_i)$ from Eq.~(\ref{nabla-eta}):
\be
   \eta_i
   =
   -
   \frac{\sqrt{\alpha}}{2}
   \int_{\theta_0}^{\theta_i}
   \frac{1}
     {(\cos\theta_0-\cos\theta)
      + \frac{\alpha}{\sin^2\theta_0}}
   \frac{d\theta}
   {{\cal R}(\theta)}.
\ee
Converting this to elliptic integrals
and using Eq.~(\ref{xi}) we get
\be
   \eta_i
   =
   -
   \frac{\sqrt{\eps\alpha}\, x_i}{(p-a)}
   -
   \frac{x_i}{|x_i|}
   \frac{\sqrt\alpha (b-a)}{\sqrt{a-c}}
     \frac{\Pi(\varphi_i,k^2\frac{p-a}{p-b},k)}{(p-a)(p-b)},
\ee
where
$p = \cos\theta_0 + \alpha/\sin^2\theta_0$.
Note that $\eta(x)$ is an odd function of $x$.

The functional determinants of $\hat A(x)$
involved in Eqs.~(\ref{eq:WL}) and (\ref{eq:II})
can be calculated numerically by introducing a proper grid $x_i$.
Then we discretize the Laplace operator in the operator $\hat A$
[Eq.~(\ref{A})], and find $\alpha_{mn}$, $\rho_{mn}$, $\eta_m$ and $\eta_n$
for each $x_i$, thus defining a finite matrix $\hat A_{ij}$.
The functional determinants in Eqs.~(\ref{eq:WL})
and (\ref{eq:II}) can then be approximated by determinants
of the matrix $\hat A_{ij}$ which should be evaluated numerically.

\section{Derivation of the operator $A$ for a wire}
\label{A:A-wire}

Substituting Eq.~(\ref{Q-W}) into (\ref{action-wire})
and expanding the action to the second order in $W$, we get:
\begin{multline}
\label{action-wire2}
  S^{(2)}
  =
  \frac{G_N}{16G_Q} \int_{-1/2}^{1/2} d x \tr
   \Bigl[
     - (\nabla W)^2
     + \left\{{\cal J},W\right\}^2
\\ {}
     - 2{\cal J}\PauliNambu_1[W,\nabla W]
     - 2\eps U\PauliNambu_3U^\dagger \PauliNambu_1 W^2
   \Bigr] ,
\end{multline}
where ${\cal J}=U\nabla U^\dagger\PauliNambu_1$.
Using the decomposition (\ref{eq:W}) of $W$ in terms
of the $d$- and $c$-modes, we rewrite $S^{(2)}$
in the form (\ref{S2-wire}), where
\begin{multline}
\label{A-SNS}
  \hat{A}_{mn}(x)=
  -\nabla^2 +\alpha_{mn} + \lambda_{mn}^2/4
\\ {}
  -i\PauliDC_2[\lambda_{mn} \nabla+(\nabla \lambda_{mn})/2]
  +\beta_{mn}\PauliDC_3
  +\gamma_{mn}\PauliDC_1.
\end{multline}
Here, $m=(\eps,a)$ and $n=(\eps',b)$ are energy and replica indices,
$\PauliDC_{i=1,2,3}$ are the Pauli matrices in the $(d,c)$-space,
$\alpha_{mn}$ is given by Eq.~(\ref{alpha}), and
\begin{subequations}
\begin{align}
&  \lambda_{mn}
  =
  -(\cos\theta_n\nabla\phi_n+\cos\theta_m\nabla\phi_m),
\\
&  \beta_{mn}
  =
  (\sin\theta_n\nabla\phi_n\sin\theta_m\nabla\phi_m-\nabla\theta_n\nabla\theta_m)/2,
\\
&  \gamma_{mn}
  =
  -(\sin\theta_n\nabla\phi_n\nabla\theta_m+\sin\theta_m\nabla\phi_m\nabla\theta_n)/2.
\end{align}
\end{subequations}
Here, ($\theta_m,\phi_m$) and ($\theta_n,\phi_n$) are the solutions
of the Usadel equations (\ref{usadeltot}) at energies $\eps$ and $\eps'$,
and phase difference $\chi_1$ or $\chi_2$, depending on the replica
indices $a$ and $b$, respectively.

In order to remove the first order derivative in Eq.~(\ref{A-SNS}),
we make a local unitary transformation in the $(d,c)$-space:
$(d,c)^T_{mn}\rightarrow {\cal V}_{mn} (d,c)^T_{mn}$, where
\begin{subequations}
\begin{gather}
\label{dc-rotation}
  {\cal V}_{mn}=\cos\frac{\zeta_n+\zeta_m}{2}
  -i\PauliDC_2\sin\frac{\zeta_n+\zeta_m}{2},
\\
  \zeta_m(x)=-\int_0^x ds\cos\theta_m(s)\nabla\phi_m(s).
\end{gather}
\end{subequations}
Such a rotation leaves $\det\hat A$ and $A_{\eps,-\eps}^{\chi\chi}$
invariant, while the operator $\hat{A}$ is transformed
to
\be
\label{A-rotation}
  \hat{\tilde A}
  =
  {\cal V}_{mn}^\dagger\hat{A}_{mn}{\cal V}_{mn},
\ee
that can be written in the form (\ref{A}) (tilde omitted)
with
\be
\label{eta-appendix}
   \eta_m =
 - \arctan \frac{\nabla\theta_m}{\sin\theta_m\nabla\phi_m}
 - \zeta_m .
\ee
Taking the derivative of (\ref{eta-appendix}), 
we come to Eq.~(\ref{nabla-eta}).

\newsavebox{\refone}
\newsavebox{\reftwo}
\sbox{\refone}{\bf\onlinecite{macedo}}
\sbox{\reftwo}{\bf\onlinecite{beenakker94}}

\section{Equivalence of Eq.~(\ref{II-short}) to the results
of Refs.~\usebox{\refone} and \usebox{\reftwo}}
\label{A:equivalence}

The result for the variance of the Josephson current,
$\var I(\chi)=\delta I^2(\chi)$, obtained
in Refs.~\onlinecite{macedo} and \onlinecite{beenakker94}
can be written in the form
\be
\label{app:var1}
  \var I(\chi)
  =
  \frac12
  \left( \frac{e\Delta}{\pi\hbar} \right)^2
  \int_0^\infty dk \,
  k(1-e^{-\pi k})\,|a(k)|^2,
\ee
\vskip -3mm
\be
\label{app:a}
  a(k)
  =
  \int_0^\infty
  \frac{dx\,\cos kx \, \sin\chi}{\cosh x\sqrt{\cosh^2x-\sin^2\frac{\chi}{2}}} .
\ee

In this Appendix we show that our expression (\ref{II-short})
for $\ccorr{I(\chi)I(\chi')}$ at $\chi'=\chi$ and zero temperature
gives the same result.

We start with rewriting Eq.~(\ref{II-short}) in the form of integrals
over $C$ and $C'$:

\makeatletter
\widetext@grid
\makeatother

\onecolumngrid


\be
\label{app:var2}
  \var I(\chi)
  =
  -\frac12 \left(\frac{e\Delta}{\pi\hbar}\right)^2
  \int_0^{\chi/2} \!\!
  \frac{dC\, \sin\chi}{\sin C\sqrt{\sin^2\frac{\chi}{2}-\sin^2C}}
\,
  \frac{dC'\, \sin\chi}{\sin C'\sqrt{\sin^2\frac{\chi}{2}-\sin^2C'}}
  \frac{\partial^2}{\partial C\partial C'}
  \ln \frac{\sin(C-C')}{C-C'} \frac{\sin(C+C')}{C+C'} .
\ee
Using Eqs.~(2.29) and (3.6) from Ref.~\onlinecite{macedo},
we rewrite the second derivative of the logarithm as
\be
  \frac{\partial^2}{\partial C\partial C'}
  \ln \frac{\sin(C-C')}{C-C'} \frac{\sin(C+C')}{C+C'}
  =
  -4 \int_0^\infty dk
  \frac{k \sinh kC \sinh kC'}{e^{\pi k}-1} ,
\ee
which allows to present Eq.~(\ref{app:var2}) in the form (\ref{app:var1})
with $\tilde a(k)$ instead of $a(k)$:
\be
\label{app:a-tilde}
  \tilde a(k)
  =
  \frac{1}{\sinh(\pi k/2)}
  \int_0^{\chi/2} \!\!
  \frac{dC\, \sinh kC \, \sin\chi}{\sin C\sqrt{\sin^2\frac{\chi}{2}-\sin^2C}} .
\ee

The equivalence between our result and the result of
Refs.~\onlinecite{macedo} and \onlinecite{beenakker94}
follows from the equality $a(k)=\tilde a(k)$.
To prove it, we extend integration in Eq.~(\ref{app:a})
to the real axis, substituting $\cos kx$ by $e^{ikx}$.
Then we deform the integration contour to the upper half-plane
and enclose all branch cuts of the square root:
$[i(\pi n+\pi/2-\chi/2),i(\pi n+\pi/2+\chi/2)]$, $n=0,1,\dots$
As a result, summation of $e^{-(n+1/2)\pi k}$ over $n$
yields $\sinh(\pi k/2)$ in the denominator, while integration
along the branch cut reproduces the integral in Eq.~(\ref{app:a-tilde}).
Thus, $a(k)=\tilde a(k)$ which establishes the equivalence between
the two results.

\section{Solution of the Usadel equation for a wire close to $T_c$}
\label{A:highT}

Close to $T_c$, the solution of the Usadel equation (\ref{usadeltot})
is given by Eq.~(\ref{eq:long-wireT}). It corresponds to the following
dependence of $\theta(x)$ and $\phi(x)$:
\begin{gather}
\label{theta-Tc}
  \sin\theta(x)
  = \frac{\sin\theta_s}{\sinh \kappa} \sqrt{
     \cosh 2 \kappa x(\cosh\kappa - \cos \chi)
     -(1-\cosh\kappa \cos \chi)
     } ,
\\
\label{phi-Tc}
  \tan \phi(x)
  =
  \frac{\tan\frac{\chi}{2}}{\tanh\frac{\kappa}{2}}\tanh\kappa x
\end{gather}
(this form is valid both for $\eps>0$ and $\eps<0$).
The function $\eta(x)$ defined in Eq.~(\ref{nabla-eta}) is given by
\be
  \tan \eta(x)
  =
  -
  \sign(\eps)
  \frac{\tanh\frac{\kappa}{2}}{\tan\frac{\chi}{2}}\tanh\kappa x .
\label{etaT}
\ee

Rigidity of Gaussian fluctuations near the saddle point is determined
by the operator
$\hat A_{\eps_1\eps_2} = -\nabla^2+|\eps_1|+|\eps_2|+\hat V_{\eps_1\eps_2}$,
where the operator $\hat V$ can be expanded in the Pauli matrices (\ref{V})
with the coefficients
\be
\label{V0}
  V^0(x)
  =
  - \frac12
  \left( \frac{\Delta}{|\epsilon_1|} \frac{\kappa_1}{\sinh\kappa_1} \right)^2
    (\cosh\kappa_1 - \cos\chi_1) \cosh2\kappa_1x
  - \frac12
  \left( \frac{\Delta}{|\epsilon_2|} \frac{\kappa_2}{\sinh\kappa_2} \right)^2
    (\cosh\kappa_2 - \cos\chi_2) \cosh2\kappa_2x ,
\ee
%
%
\begin{multline}
\label{V1}
  V^1(x)
  =
  -
  2
  \frac{\Delta}{|\epsilon_1|} \frac{\kappa_1}{\sinh\kappa_1}
  \frac{\Delta}{|\epsilon_2|} \frac{\kappa_2}{\sinh\kappa_2}
  \Bigr(
    \sign(\eps_1)
    \cos\frac{\chi_1}{2} \sin\frac{\chi_2}{2}
    \sinh\frac{\kappa_1}{2} \cosh\frac{\kappa_2}{2}
    \sinh\kappa_1 x \cosh\kappa_2 x
\\ {}
  + \sign(\eps_2)
    \sin\frac{\chi_1}{2} \cos\frac{\chi_2}{2}
    \cosh\frac{\kappa_1}{2} \sinh\frac{\kappa_2}{2}
    \cosh\kappa_1 x \sinh\kappa_2 x
  \Bigr) ,
\end{multline}
%
%
\begin{multline}
\label{V3}
  V^3(x)
  =
  2
  \frac{\Delta}{|\epsilon_1|} \frac{\kappa_1}{\sinh\kappa_1}
  \frac{\Delta}{|\epsilon_2|} \frac{\kappa_2}{\sinh\kappa_2}
  \Bigl(
    \sin\frac{\chi_1}{2} \sin\frac{\chi_2}{2}
    \cosh\frac{\kappa_1}{2} \cosh\frac{\kappa_2}{2}
    \cosh\kappa_1 x \cosh\kappa_2 x
\\ {}
  - \sign(\eps_1\eps_2)
    \cos\frac{\chi_1}{2} \cos\frac{\chi_2}{2}
    \sinh\frac{\kappa_1}{2} \sinh\frac{\kappa_2}{2}
    \sinh\kappa_1 x \sinh\kappa_2 x
  \Bigr) .
\end{multline}

The supercurrent correlation function (\ref{TII}) involves
\be
\label{R}
  R_{\eps_1\eps_2}
  =
  \Tr (G\hat V)^2
  =
  2 \int_{-1/2}^{1/2} dx \, dy \,
  G_{\eps_1\eps_2}^2(x,y)
  \sum_{i=0,1,3} V^i(x) V^i(y) .
\ee
Since current fluctuations are determined
by $\sum_{\eps_1,\eps_2}R_{\eps_1\eps_2}$,
it is convenient to symmetrize $R$ by introducing
\be
  \tilde R_{\eps_1,\eps_2}
  =
  \frac14
  \left[
    R_{\eps_1,\eps_2}+R_{-\eps_1,\eps_2}+R_{\eps_1,-\eps_2}+R_{-\eps_1,-\eps_2}
  \right] ,
\ee
such that
$\sum_{\eps_1,\eps_2}R_{\eps_1\eps_2}
= \sum_{\eps_1,\eps_2} \tilde R_{\eps_1\eps_2}$.

Substituting Eqs.~(\ref{V0})--(\ref{V3}) into Eq.~(\ref{R}),
and taking the derivatives with respect to $\chi_1$ and $\chi_2$
we get
\be
  \frac{\partial^2}{\partial\chi_1\partial\chi_2}
  \tilde R_{\eps_1\eps_2}
  =
  \left( \frac{\Delta}{|\epsilon_1|} \frac{\kappa_1}{\sinh\kappa_1} \right)^2
  \left( \frac{\Delta}{|\epsilon_2|} \frac{\kappa_2}{\sinh\kappa_2} \right)^2
  \Upsilon_{\eps_1\eps_2}
  \sin\chi_1\sin\chi_2 ,
\ee
resulting in Eq.~(\ref{eq:mesoT}).
Here the function $\Upsilon_{\eps_1\eps_2}$ is defined by the
double integral:
\begin{multline}
  \Upsilon_{\eps_1\eps_2}
  =
  \int_{-1/2}^{1/2} dx \, dy \, G_{\eps_1\eps_2}(x,y)^2
  \Bigl[ \cosh2\kappa_1 x \cosh2 \kappa_2y
\\ {}
  + 2\cosh^2\frac{\kappa_1}{2}\cosh^2\frac{\kappa_2}{2}
    \cosh\kappa_1 x \cosh \kappa_2 x\cosh\kappa_1 y \cosh \kappa_2 y
  + 2\sinh^2\frac{\kappa_1}{2}\sinh^2\frac{\kappa_2}{2}
    \sinh\kappa_1 x \sinh \kappa_2 x\sinh\kappa_1 y \sinh \kappa_2 y
\\ {}
  - 2\cosh^2\frac{\kappa_1}{2}\sinh^2\frac{\kappa_2}{2}
    \cosh\kappa_1 x \sinh\kappa_2 x\cosh\kappa_1 y \sinh \kappa_2 y
  - 2\sinh^2\frac{\kappa_1}{2}\cosh^2\frac{\kappa_2}{2}
    \sinh\kappa_1 x \cosh \kappa_2 x\sinh\kappa_1 y \cosh \kappa_2 y \Bigr]
  .
\label{upsilon}
\end{multline}
In principle, these integrals can be calculated in a closed form.
However, the resulting expression is too complicated and we leave
the integrals unevaluated.


\twocolumngrid


\begin{thebibliography}{99}
\bibitem{AS-comment}
     For this reason, the theory of Ref.~\onlinecite{altshuler87}
     does not allow determining correctly the average Josephson current.
\bibitem{A-rotated}
We use the same notation for the initial matrix $\hat A$ and the rotated
matrix ${\cal V}^\dagger \hat{A} {\cal V}$ since they give identical
WL corrections and mesoscopic fluctuations.
\end{thebibliography}
\end{document}